\documentclass[a4paper,11pt]{article}
\pdfoutput=1
\usepackage[utf8]{inputenc}
\usepackage{jcappub}
\usepackage[T1]{fontenc}

\usepackage[mathscr]{eucal}
\usepackage[Symbolsmallscale]{upgreek}

\newcommand{\Tdec}{T_\text{dec}}
\newcommand{\Tend}{T_\text{end}}
\newcommand{\fend}{f_\text{end}}
\newcommand{\kend}{k_\text{end}}
\newcommand{\Teq}{T_\text{eq}}
\newcommand{\feq}{f_\text{eq}}
\newcommand{\keq}{k_\text{eq}}
\newcommand{\Tstart}{T_\text{start}}
\newcommand{\Tmax}{T_\text{max}}
\newcommand{\aend}{a_\text{end}}
\newcommand{\aeq}{a_\text{eq}}
\newcommand{\adec}{a_\text{dec}}
\newcommand{\astart}{a_\text{start}}
\newcommand{\amax}{a_\text{max}}
\newcommand{\Thc}{T_\text{hc}}
\newcommand{\ahc}{a_\text{hc}}

\newcommand{\tHmax}{\tilde H_\text{max}}
\newcommand{\gs}{g_\star}
\newcommand{\hs}{h_\star}
\newcommand{\op}{\omega_\phi}
\newcommand{\rp}{\rho_\phi}
\newcommand{\rR}{\rho_R}
\newcommand{\OGW}{\Omega_\text{GW}}
\newcommand{\ximin}{\xi_\text{min}}

\title{Primordial Gravitational Waves\\in Nonstandard Cosmologies}

\author[a]{Nicolás Bernal}
\author[b]{and Fazlollah Hajkarim}

\affiliation[a]{Centro de Investigaciones, Universidad Antonio Nariño,\\
Carrera 3 Este \# 47A-15, Bogotá, Colombia}
\affiliation[b]{Institut für Theoretische Physik, Goethe Universität,\\
Max von Laue Straße 1, D-60438 Frankfurt, Germany}

\emailAdd{nicolas.bernal@uan.edu.co}
\emailAdd{hajkarim@th.physik.uni-frankfurt.de}

\abstract{Assuming that inflation is followed by a phase where the energy density of the Universe is dominated by a component with a general equation of state, we evaluate the spectrum of primordial gravitational waves induced in the post-inflationary Universe.
We show that if the energy density of the Universe is dominated by a component $\phi$ before Big Bang nucleosynthesis, its equation of state could be constrained by  gravitational wave experiments depending on the ratio of energy densities of $\phi$ and radiation, and also the temperature at the end of the $\phi$ dominated era.
Also, we discuss the impact of scale dependence of tensor modes on the primordial gravitational wave spectrum during the $\phi$-domination.
These models are motivated by beyond Standard Model physics and scenarios for non-thermal production of dark matter in the early Universe.
We also constrain the parameter space of the tensor spectral index and the tensor-to-scalar ratio, using the experimental limits from gravitational wave experiments.}

\begin{document}

\begin{flushright}
PI/UAN-2019-648FT
\end{flushright}

\maketitle

\section{Introduction}

The recent observations of Gravitational Waves (GW) by LIGO and Virgo~\cite{Abbott:2016blz, Abbott:2016nmj, Abbott:2017vtc, Abbott:2017oio, TheLIGOScientific:2017qsa, Abbott:2017gyy} paved the way to observe the Universe with new methods not based on electromagnetic radiation.
Until now our knowledge about the early Universe cosmology was limited by electromagnetic waves back to last scattering surface of photons, some possible effects of inflationary scenario on the cosmic microwave background, and the abundance of light elements from Big Bang Nucleosynthesis (BBN).
Although the current GW detectors are only sensitive to strong astrophysical events such as merging black holes or neutron stars, future experiments are expected to detect much weaker signatures produced in the early Universe~\cite{Guzzetti:2016mkm, Caprini:2018mtu}.
Several space-borne interferometers such as the proposed ground-based Einstein Telescope (ET)~\cite{Sathyaprakash:2012jk}, the planned space-based LISA~\cite{Audley:2017drz} interferometer, the proposed successor experiments BBO~\cite{Crowder:2005nr}, (B-)~DECIGO~\cite{Seto:2001qf, Sato:2017dkf}, as well as the SKA~\cite{Janssen:2014dka} telescope are planned to be operational in the future with the aim of detecting the primordial GW (PGW) background and the effect of possible cosmic phase transitions on it.

The existence of a PGW background is one of the most crucial predictions of the inflationary scenario of the early Universe~\cite{Grishchuk:1974ny, Starobinsky:1979ty}.
The spectrum of the inflationary GWs that could be observed today depends on two main factors: one is the power spectrum of primordial tensor perturbations generated during inflation, and the other is the expansion rate of the Universe from the end of inflation until today.
The former defines the initial magnitude of the GW signature, and it is directly associated with the detailed properties of inflationary models.
The latter describes how the density of the PGWs has been diluted in subsequent stages of the cosmic expansion.
Since the amplitude and polarization of PGWs can be modified by non-standard cosmological scenarios, there is a possibility to extract information about the early Universe using GW experiments~\cite{Kuroyanagi:2018csn}. 

On the one hand, concerning the PGW spectrum, current CMB measurements do not have the ability to constrain the amplitude $A_T$ nor the tensor spectral index $n_T$.
The measurement of the tensor-to-scalar ratio $r$ is still compatible with zero, and for low enough $r$, practically any value of $n_T$ is still acceptable.
For this reason, the constraints on $n_T$ depend on the chosen prior on $r$.
This situation will change when a positive detection of a non-zero tensor amplitude is obtained from primordial $B$-modes~\cite{Bouchet:2011ck, Cabass:2015jwe}.

On the other hand, several effects like the decoupling of neutrinos or the variation of Standard Model (SM) relativistic degrees of freedom, alter the nature of the GW spectrum during its propagation~\cite{1982ApJ...257..456V, Rebhan:1994zw, Schwarz:1997gv, Seto:2003kc, Weinberg:2003ur, Bashinsky:2005tv, Dicus:2005rh, Boyle:2005se, Watanabe:2006qe, Kuroyanagi:2008ye, Mangilli:2008bw, Stefanek:2012hj, Dent:2013asa, Baym:2017xvh, Caldwell:2018giq}.
However, one can imagine that instead of being dominated by radiation over its early phase (i.e. the standard cosmological scenario), the evolution of the Universe could have been driven by a matter, or in general by a component $\phi$ with a general equation of state $\op$.
In fact, there are no fundamental reasons to assume that the Universe was radiation-dominated prior to BBN\footnote{For studies on baryogenesis with a low reheating temperature or during an early matter-dominated phase, see refs.~\cite{Davidson:2000dw, Giudice:2000ex, Allahverdi:2010im, Beniwal:2017eik, Allahverdi:2017edd, Dutta:2018zkg} and~\cite{Bernal:2017zvx}, respectively.} at $t\sim 1$~s.
Studying what consequences such a non-standard era can have on observational properties of GW is hence worthwhile.
In particular, GW in scenarios with an early matter era have received particular attention~\cite{Assadullahi:2009nf, Durrer:2011bi, Alabidi:2013wtp, DEramo:2019tit, Inomata:2019zqy, Inomata:2019ivs}.
Additionally, let us note that production of dark matter in scenarios with non-standard expansion phases has recently gained increasing interest~\cite{Kamionkowski:1990ni, Salati:2002md, Comelli:2003cv, Rosati:2003yw, Gelmini:2006pw, Gelmini:2006pq, Arbey:2008kv, Gelmini:2008sh, Arbey:2009gt, Visinelli:2009kt, Co:2015pka, Berlin:2016vnh, Tenkanen:2016jic, Dror:2016rxc, Berlin:2016gtr, Dutta:2016htz, DEramo:2017gpl, Hamdan:2017psw, Visinelli:2017qga, Dror:2017gjq, Drees:2017iod, DEramo:2017ecx, Bernal:2018ins, Hardy:2018bph, Maity:2018exj, Hambye:2018qjv, Bernal:2018kcw, Arbey:2018uho, Nelson:2018via, Drees:2018dsj, Betancur:2018xtj, Arbuzova:2018apk, Maldonado:2019qmp, Arias:2019,Heurtier:2019eou,Maity:2018dgy}.

Previous works have already investigated the degree to which the thermal history and the the early Universe equation of state affect the propagation of GW~\cite{Sahni:1990tx,Giovannini:1998bp, Giovannini:1999bh, Riazuelo:2000fc, Sahni:2001qp, Tashiro:2003qp, Seto:2003kc, Boyle:2005se, Watanabe:2006qe, Saikawa:2018rcs, Caldwell:2018giq, Ramberg:2019dgi,Opferkuch:2019zbd,Lozanov:2017hjm,Lozanov:2016hid,Nunes:2018zot}.
Also how the pre-BBN Universe could be probed with GW from cosmic strings~\cite{Cui:2017ufi, Cui:2018rwi} or from gravitational reheating~\cite{Artymowski:2017pua}.
Instead, here we perform a full numerical evaluation of the PGW spectrum, solving properly the equations for the background energy density, and taking special care of the variation of the SM degrees of freedom.
For different PGW spectra and varying the possible thermal histories of the Universe, we explore the capabilities of current and future GW detectors to probe the GW background.

The rest of this paper is organized as follows. 
In section~\ref{pgw} we revisit the set of differential equations that govern the tensor perturbations.
Then we compute the spectrum of GW in the standard radiation dominated period.
In section~\ref{sec:non} we introduce our setup for non-standard cosmologies to include possible equations of state for the fluid $\phi$ and their impact on the Hubble expansion rate and the thermal history of the Universe. 
Section~\ref{sec:pgw-non} is devoted to the calculation of relic GW spectrum in case of a scale invariant power spectrum and assuming $\phi$ dominated era.
We also perform a scan over the parameter space of possible equation-of-states and ratio of densities for radiation and the non-standard fluid.
The effect of scale dependence on the spectrum of GW on the parameter space is also studied.
Finally, we conclude and summarize our results in section~\ref{sec:conclusion}.

\section{Primordial Gravitational Wave Spectrum}
\label{pgw}

GWs are represented by spatial metric perturbations that satisfy the transverse-traceless conditions: $\partial^ih_{ij}=0$ and $h_i^i=0$.
The evolution of GWs is described by the linearized Einstein equation
\begin{equation}\label{eq:gw1}
    \ddot h_{ij}+3H\,\dot h_{ij}-\frac{\nabla^2}{a^2}h_{ij}=16\pi\,G\,\Uppi_{ij}^{TT},
\end{equation}
where the dots correspond to derivatives with respect to the cosmic time $t$, and $G$ is the Newton's constant.
$\Uppi_{ij}^{TT}$ is the transverse-traceless part of the anisotropic stress tensor $\Uppi_{ij}$
\begin{equation}
    \Uppi_{ij}=\frac{T_{ij}-p\,g_{ij}}{a^2}\,,
\end{equation}
where $T_{ij}$ is the stress-energy tensor, $g_{ij}$ is the metric tensor and $p$ the background pressure.
The spatial metric perturbations can be decomposed into their Fourier modes
\begin{equation}
    h_{ij}(t,\vec x)=\sum_\lambda\int\frac{d^3k}{(2\pi)^3}\,h^\lambda(t,\vec k)\,\upepsilon_{ij}^\lambda(\vec k)\,e^{i\,\vec k\cdot\vec x},
\end{equation}
where $\lambda=+,\,\times$ corresponds to the two independent polarization states, and $\upepsilon_{ij}^\lambda(\vec k)$ are the spin-2 polarization tensors satisfying the normalization conditions $\sum_{ij}\upepsilon_{ij}^\lambda\,{\upepsilon_{ij}^{\lambda'}}^*=2\delta^{\lambda\lambda'}$.
Equation~\eqref{eq:gw1} can therefore be rewritten as
\begin{equation}\label{eq:gw2}
    \ddot h_{\vec k}^\lambda+3H\,\dot h_{\vec k}^\lambda+\frac{k^2}{a^2}h_{\vec k}^\lambda=16\pi\,G\,\Uppi_{\vec k}^\lambda\,,
\end{equation}
where $h_{\vec k}^\lambda(t)\equiv h^\lambda(t,\vec k)$. In the rest of this paper we consider the RHS of the above equation to be zero so it does not enhance the primordial tensor perturbations.
However, in general it is finite, for example when one considers the effect of damping of photons and neutrinos at low frequencies or the impact of scalar perturbations which can act as a source for tensor perturbations~\cite{Caprini:2018mtu, Saikawa:2018rcs, Scomparin:2019ziw}.
We do not consider such effects in this paper.
The solution of eq.~\eqref{eq:gw2} can be expressed as
\begin{equation}\label{eq:trans}
    h_{\vec k}^\lambda=h_{\vec k,\,\text{prim}}^\lambda\, X(t,k)\,,
\end{equation}
where $h_{\vec k,\,\text{prim}}^\lambda$ represents the amplitude of the primordial tensor perturbations and $X(t,k)$ is the transfer function, normalized such that $X(t,k)=1$ for $k\ll a\,H$.

The energy density of the relic GWs is given by
\begin{equation}
    \rho_\text{GW}(t)=\frac{1}{16\pi\,G}\sum_\lambda\int\frac{d^3k}{(2\pi)^3}\left|\dot h_{\vec k}^\lambda\right|^2.
\end{equation}
The primordial gravitational wave spectrum is calculated following refs.~\cite{Watanabe:2006qe,Saikawa:2018rcs}:
\begin{equation}
    \OGW(t,k)=\frac{1}{\rho_c(t)}\,\frac{d\rho_\text{GW}(t,k)}{d\ln k},
\end{equation}
where $\rho_c$ is the critical energy density of the Universe.
This spectrum can be rewritten using eq.~\eqref{eq:trans} as
\begin{equation}
    \OGW(\eta,k)=\frac{1}{12\,a^2(\eta)\,H^2(\eta)}\mathscr{P}_T(k)\,\left[X'(\eta,\,k)\right]^2,
\end{equation}
where the prime represents the derivative with respect to the conformal time $\eta$.
The primordial tensor power spectrum $\mathscr{P}_T(k)$ is determined by the Hubble parameter at the time when the corresponding mode crosses the horizon during inflation, at $k=a\,H$,
\begin{equation}
    \mathscr{P}_T(k)\equiv\frac{k^3}{\pi^2}\sum_\lambda \left|h^\lambda_{\vec{k},\,\text{prim}}\right|^2=\left.\frac{2\,H^2}{\pi^2M_{Pl}^2}\right|_{k=a\,H},
\end{equation}
where $M_{Pl}=2.435\times 10^{18}$~GeV is the reduced Planck mass. 

The transfer function is found by numerically solving the equation
\begin{equation}\label{eq:gwode}
    \frac{d^2X(u)}{du^2}+\frac{2}{a(u)}\,\frac{da(u)}{du}\,\frac{dX(u)}{du}+X(u)=0,
\end{equation}
where $u\equiv k\,\eta$.
Again, although the shear of the cosmic fluid gives rise to some important effects~\cite{Watanabe:2006qe}, we ignore its possible contribution by setting the RHS of eq.~\eqref{eq:gwode} to zero, for frequencies beyond $\sim 10^{-10}$~Hz.
In the frequency range between $\sim 10^{-16}$ and $\sim 10^{-10}$~Hz, the damping effect due to the free-streaming neutrinos reduce the amplitude of GW by $\sim 35\%$~\cite{1982ApJ...257..456V, Rebhan:1994zw, Weinberg:2003ur}, which is not interesting for us in this paper.
The initial conditions are specified as
\begin{equation}
    X(0)=1,\qquad\qquad\frac{dX}{du}(0)=0.
\end{equation}
The wave equation~\eqref{eq:gwode} is solved up to some finite time $u^\prime$ after horizon crossing; after that we extrapolate the solution until the present time by using the WKB solution,
\begin{equation}
    X(u)=\frac{A}{a(u)}\,\sin(u+\delta),
\end{equation}
where $A$ and $\delta$ are fixed such that $X$ and $dX/du$ match the numerical solution at $u=u^\prime$.

\begin{figure}[t]
\begin{center}
\includegraphics[height=0.55\textwidth]{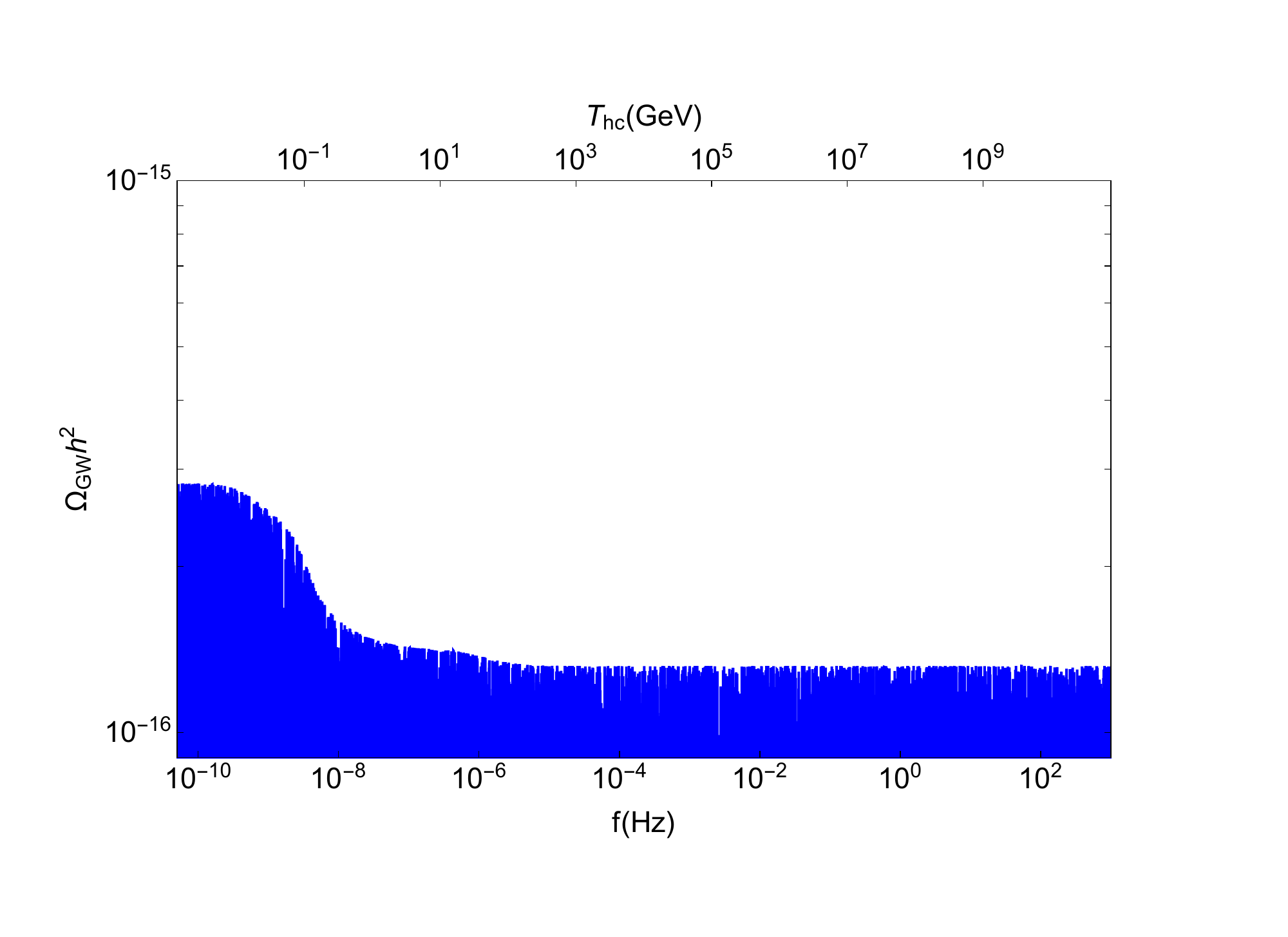}
\caption{The spectrum of inflationary GWs ($\OGW h^2$) as a function of the frequency $f$, for the standard cosmological scenario. Here we fix the inflationary scale as $V_\text{inf}^{1/4}=1.5\times 10^{16}$~GeV and assume a primordial scale invariant spectrum, i.e. $n_T=0$. We also show the temperature $\Thc$ at which the corresponding mode reenters the horizon.
}
\label{fig:OGWSM}
\end{center}
\end{figure}
Figure~\ref{fig:OGWSM} shows the result of the numerical integration of the spectrum of inflationary GWs as a function of the frequency $f\equiv k/(2\pi)$, for the standard cosmological scenario.
We have fixed the inflationary scale as $V_\text{inf}^{1/4}=1.5\times 10^{16}$~GeV,\footnote{This value  comes from the fact that at the end of inflation after $60$ $e$-folds the value of Hubble parameter is $H_\text{inf}\sim 10^{-5} M_{Pl}$~\cite{Liddle:1993ch}.} i.e. $\mathscr{P}_T(k)=\frac{2}{3\pi^2}\frac{V_\text{inf}}{M_{Pl}^4}$ and assume a primordial scale invariant scenario, i.e. $n_T=0$ (also  see Sec.~\ref{sec:tentilt}).
We also show the temperature $\Thc$ at which the corresponding mode reenters the horizon.
The oscillatory behavior is a genuine feature of inflationary GWs.
The decrease in the spectrum between $\sim 10^{-9}$ and $\sim 10^{-8}$~Hz corresponds to the variation of the relativistic degrees of freedom due to the QCD smooth crossover transition, where we used the SM equation of state from ref.~\cite{Drees:2015exa}.
Moreover, it was shown that using a different lattice QCD equation of state for the calculation of the SM equation of state only affects the predicted PGW spectrum at the order of a few percent~\cite{Schettler:2010dp, Hajkarim:2019csy}.
The dependence on the number of relativistic degrees of freedom $\gs$ and $h_\star$ that contribute to the SM energy density and the SM entropy density respectively is~\cite{Saikawa:2018rcs}
\begin{equation}\label{gwrelic-smdof}
\OGW(\eta_0,\,k)\approx \frac{\Omega_{\gamma}(T_0)}{48}\,\gs(\Thc)\,\left[\frac{\hs(T_0)}{\hs(\Thc)}\right]^{4/3}\mathscr{P}_T(k).
\end{equation}
Here $\Omega_\gamma$ corresponds to the photon relic density, $T_0$ and $\Thc$ correspond to today's and horizon crossings temperatures, respectively.

\section{Non-standard Cosmologies}
\label{sec:non}

We assume that for some period of the early Universe, the total energy density was dominated by a component $\rp$ with an equation of state parameter $\op$, where $\op\equiv p_\phi/\rp$, with $p_\phi$ the pressure of the dominant component.
We assume that this component decays solely into SM radiation with a rate $\Gamma_\phi$.
In the early Universe, the evolution of the energy density $\rp$ and the SM entropy density $s_R$ are governed by the system of coupled Boltzmann equations
\begin{align}
\frac{d\rp}{dt}+3(1+\op)\,H\,\rp&=-\Gamma_\phi\,\rp\,,\label{eq:cosmo2} \\
\frac{ds_R}{dt}+3\,H\,s_R&=+\frac{\Gamma_\phi\,\rp}{T}\,.\label{eq:cosmo3}
\end{align}
Under the assumption that the SM plasma maintains internal equilibrium at all times in the early Universe, the temperature dependence of the SM energy density $\rR$ can be obtained from
\begin{equation}\label{eq:rhoR}
    \rR(T)=\frac{\pi^2}{30}\,\gs(T)\,T^4.
\end{equation}
Equation~\eqref{eq:cosmo3} plays an important role in tracking properly the evolution of the photon's temperature $T$, via the SM entropy density $s_R$
\begin{equation}\label{eq:entropy}
    s_R(T)=\frac{\rR+p_{R}}{T}=\frac{2\pi^2}{45}\,h_\star(T)\,T^3,
\end{equation}
where $\gs(T)$ and $h_{\star}(T)$ correspond to the effective number of relativistic degrees of freedom for the SM energy and entropy densities~\cite{Drees:2015exa}. 
The Hubble expansion rate $H$ is defined by 
\begin{equation}
    H^2=\frac{\rp+\rR+\rho_m+\rho_\Lambda}{3\,M_{Pl}^2}\,,\label{eq:hubble}
\end{equation}
where $\rho_m$ and $\rho_\Lambda$, corresponding to the matter and cosmological constant energy densities respectively, are subdominant before the matter-radiation equality. 

Using entropy conservation in standard cosmology we can compute the evolution of the temperature with respect to the scale factor using
\begin{equation}
\frac{dT}{da} =\left[1+\frac{T}{3h_\star}\frac{dh_\star}{dT}\right]^{-1}
\left[-\frac{T}{a}\right].\label{eq:cosmotemp0}
\end{equation}
However, once we assume a period of $\phi$ domination which decays to radiation the entropy is not conserved anymore and from eqs.~\eqref{eq:cosmo3} and~\eqref{eq:entropy} one has
\begin{equation}
\frac{dT}{da} =\left[1+\frac{T}{3h_\star}\frac{dh_\star}{dT}\right]^{-1}
\left[-\frac{T}{a}+\frac{\Gamma_\phi\,\rp}{3\,H\,s\,a}\right].\label{eq:cosmotemp}
\end{equation}

The approximate temperature $\Tdec$ at which $\phi$ decays is fixed by the total decay width $\Gamma_\phi$ as
\begin{equation}\label{eq:Tend}
\Tdec^4=\frac{90}{\pi^2\,\gs(\Tdec)}\,M_{Pl}^2\,\Gamma_\phi^2.
\end{equation}
For having a successful BBN, that temperature has to be $\Tdec\gtrsim T_\text{BBN}\sim 4$~MeV~\cite{Kawasaki:2000en, Hannestad:2004px, Ichikawa:2005vw, DeBernardis:2008zz, deSalas:2015glj}.
To present the maximal effect that a non-standard expansion phase can have on the GW spectrum, we choose $\Tdec=10$~MeV, which is close to the BBN bound.\footnote{Let us note that for $\op>1/3$, $\rp$ gets dissolved faster than radiation.
If $\rp\ll\rR$ at $T_\text{dec}$, $\Gamma_\phi$ could effectively be taken to zero.}
However, the results can be easily generalized to higher values of $\Tdec$.
The scale factor at the moment when $\phi$ decays is denoted by $\adec$. 

The initial condition used to compute the evolution of Boltzmann equations is 
\begin{equation}
\xi\equiv\left.\frac{\rp}{\rR}\right|_{T=\Tmax}
\end{equation}
with $\Tmax=10^{14}$~GeV.
In complete inflationary scenarios $\xi$ is a theoretical prediction and not an input parameter.
Let us emphasize that the choice of $\Tmax$ is not physical, and therefore it should be taken as a simple pivot scale from which we start to solve the Boltzmann equations, and not as the maximal temperature reached by the thermal bath.
The total energy density at $T=\Tmax$ is the sum of radiation and $\phi$, so that $\rho(\Tmax)=(\rR+\rp)|_{T=\Tmax}=\rR(\Tmax)\times(1+\xi)$.
We solve eqs.~\eqref{eq:cosmo2} and~\eqref{eq:cosmotemp} numerically to find the evolution of temperature with respect to scale factor in a non-standard cosmological scenario.
The scale factor $a$ as a function of conformal time $\eta$ can then be used as input for eq.~\eqref{eq:gwode} to calculate the spectrum of GW background under a $\phi$ dominated era.

\begin{figure}[t]
\begin{center}
\includegraphics[height=0.35\textwidth]{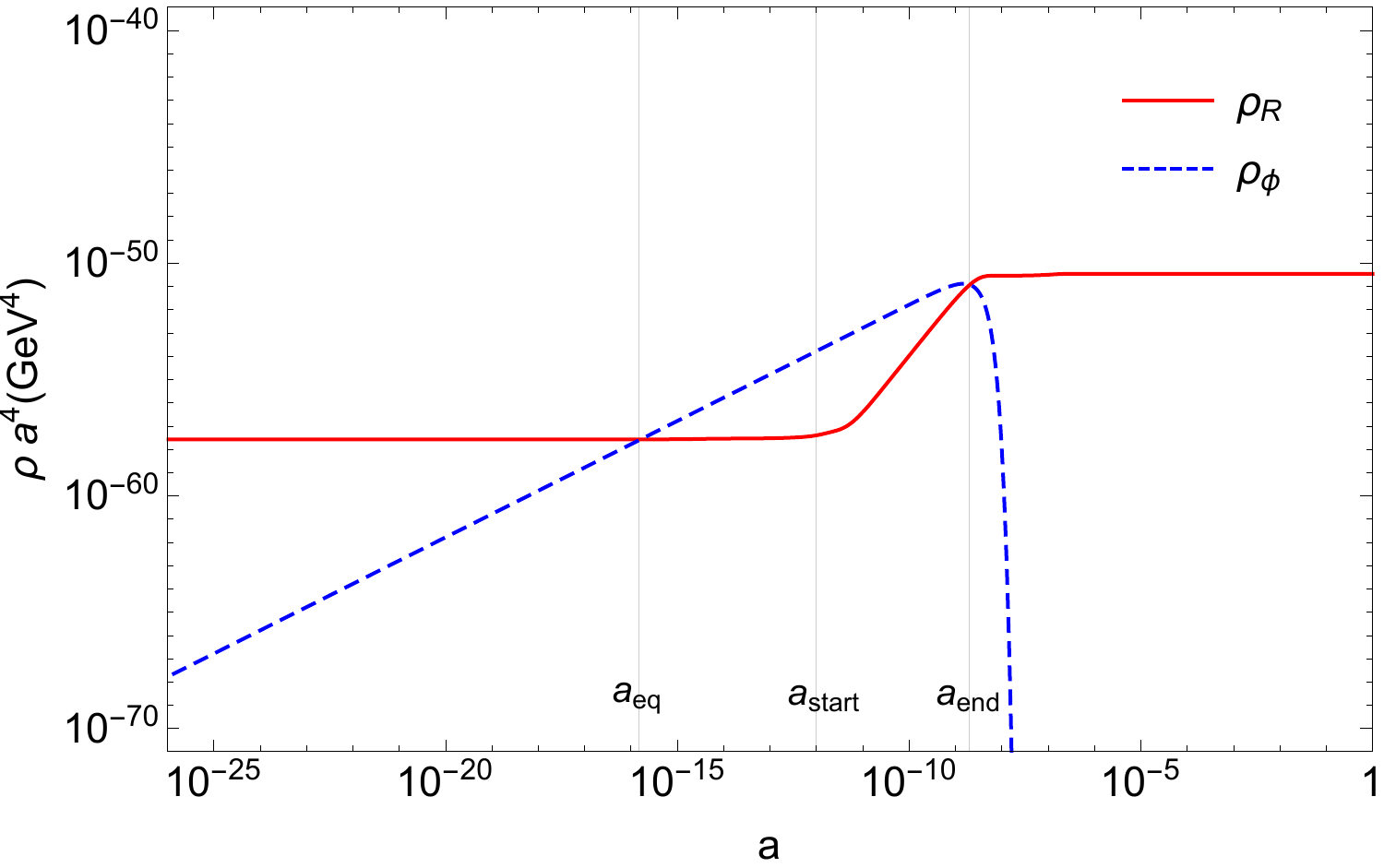}
\includegraphics[height=0.35\textwidth]{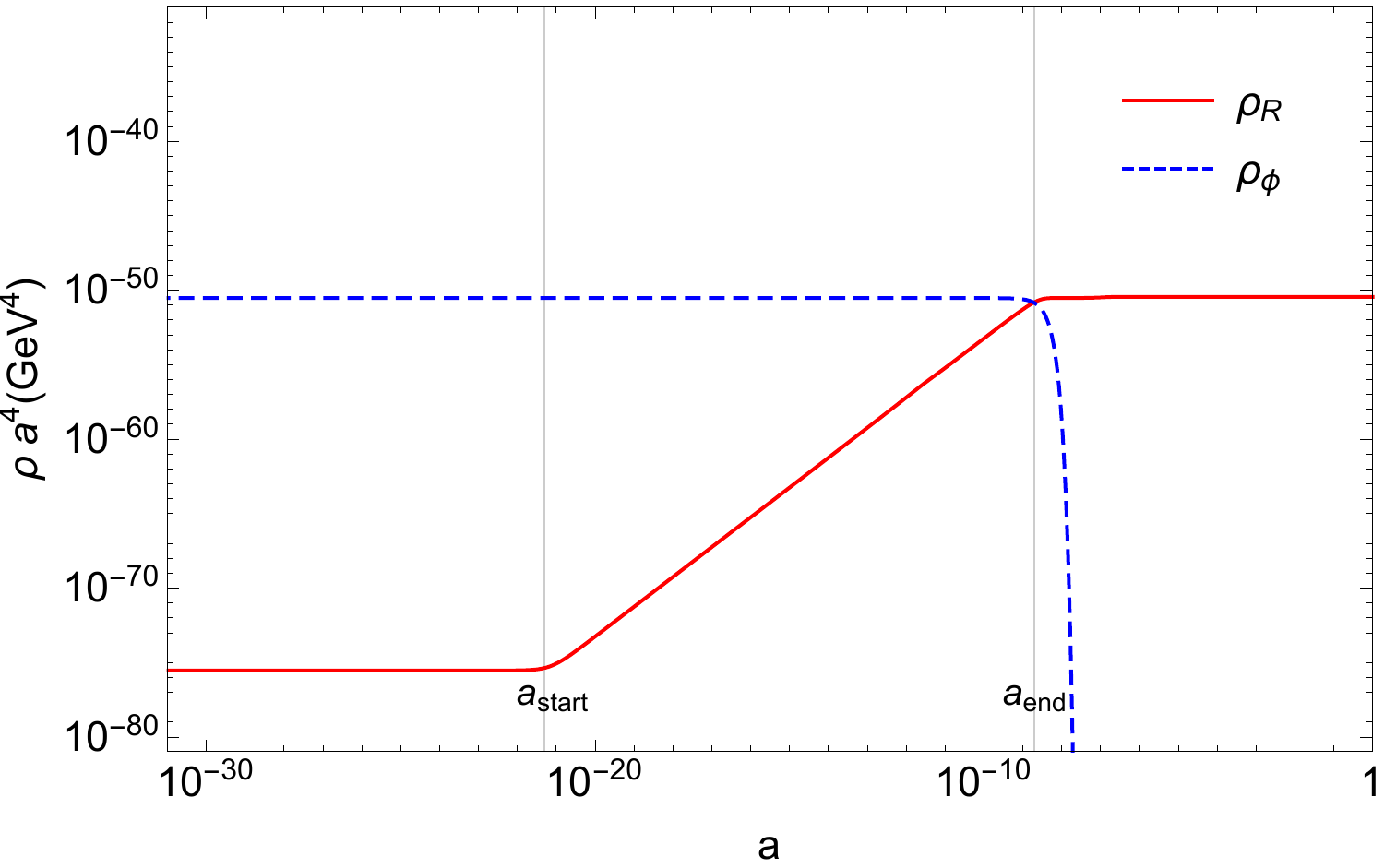}
\includegraphics[height=0.35\textwidth]{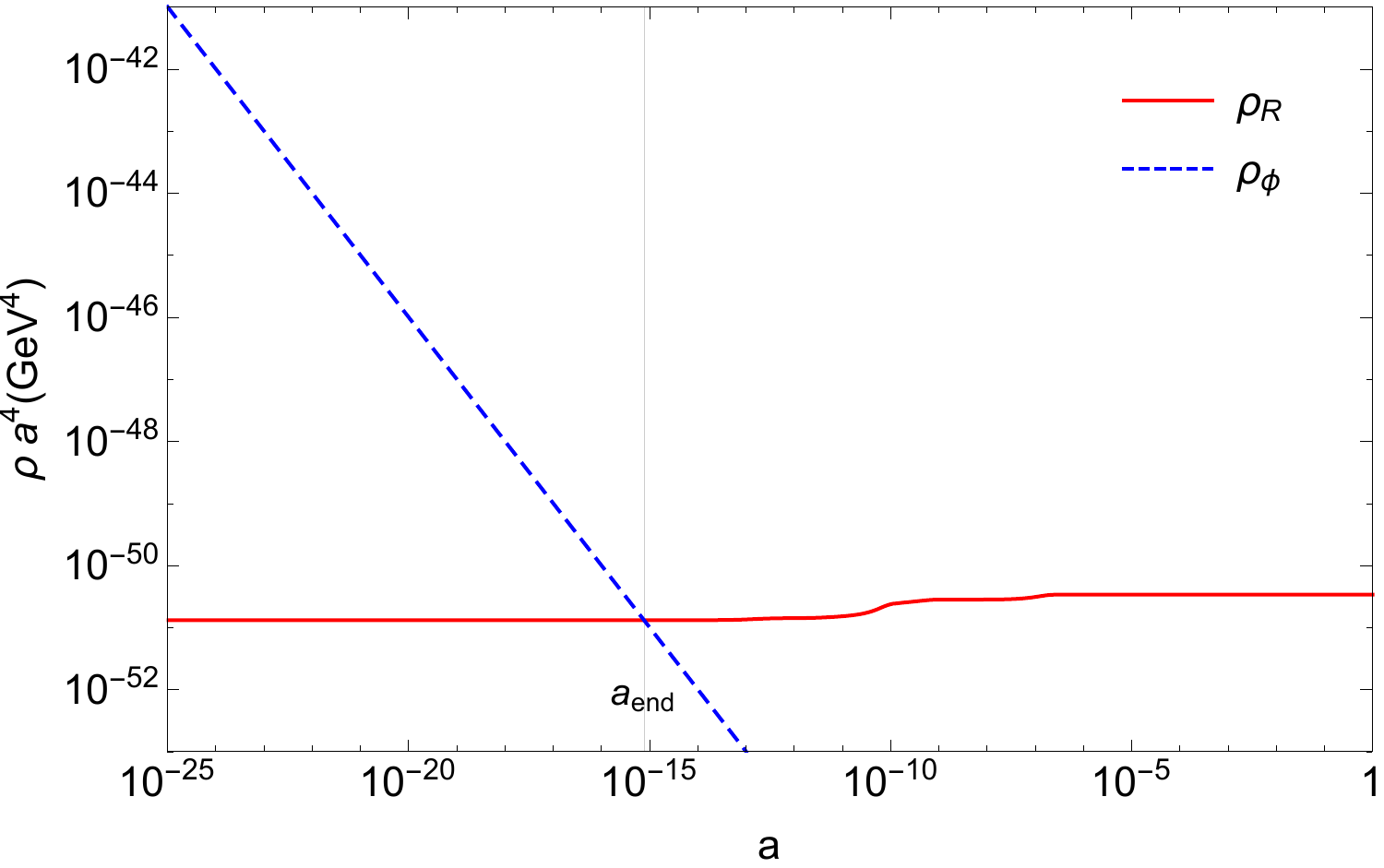}
\caption{Example of the evolution of the energy densities $\rR$ and $\rp$ as a function of the scale factor $a$, for $\op=0$ and $\xi=10^{-11}$ (upper panel), $\op=1/3$ and $\xi=10^{25}$ (central panel), and $\op=2/3$ and $\xi=10^{10}$ (lower panel).
We have chosen $\Tdec=10$~MeV.
These benchmark points are the same used in fig.~\ref{fig:OGW} and are shown in the upper left panel of fig.~\ref{fig:treh-vinf}. 
The vertical gray lines corresponding to $a=\aeq$, $\astart$ and $\aend$ are overlaid.
}
\label{fig:rho}
\end{center}
\end{figure}

As an example, fig.~\ref{fig:rho} shows the evolution of the energy densities $\rR$ and $\rp$ as a function of the scale factor $a$, for $\op=0$ and $\xi=10^{-11}$ (upper panel), $\op=1/3$ and $\xi=10^{25}$ (central panel), and $\op=2/3$ and $\xi=10^{10}$ (lower panel).
We have chosen $\Tdec=10$~MeV.
In fig.~\ref{fig:rho} the value of radiation energy density at $T=T_0$ (i.e. $a=a_0=1$) matches the CMB energy density.
If one ignores the variation of the number of relativistic degrees of freedom $\gs$ and $\hs$, one has that $\rp(a)\propto a^{-3(1+\op)}$ until it decays, and
\begin{equation}\label{eq:rhoRa}
    \rR(a)\propto
    \begin{cases}
    a^{-4}\hspace{1.82cm}\text{ for }\hspace{1.4cm} a\ll \astart,\\
    a^{-\frac32(1+\op)}\qquad\text{ for } \astart\ll a\ll \adec,\\
    a^{-4}\hspace{1.82cm}\text{ for }\hspace{0.1cm} \adec\ll a,
    \end{cases}
\end{equation}
which by using eq.~\eqref{eq:rhoR} implies that
\begin{equation}\label{eq:T}
    T(a)\propto
    \begin{cases}
    a^{-1}\hspace{1.82cm}\text{ for }\hspace{1.4cm} a\ll \astart,\\
    a^{-\frac38(1+\op)}\qquad\text{ for } \astart\ll a\ll \adec,\\
    a^{-1}\hspace{1.82cm}\text{ for }\hspace{0.1cm} \adec\ll a.
    \end{cases}
\end{equation}
Additionally, let us define $\Teq\equiv T(a=\aeq)$, $\Tstart\equiv T(a=\astart)$ and $\Tend\equiv T(a=\aend)$ (see appendix~\ref{sec:appendix}).
$\Tdec$ is properly defined in eq.~\eqref{eq:Tend}.
$\Teq$ corresponds to the temperature at which $\rp=\rR$, well before $\phi$ decays, in the case where $\op<1/3$.
In fig.~\ref{fig:rho} the vertical gray lines corresponding to $a=\aeq$, $\astart$ and $\aend$ are overlaid.
Moreover, in fig.~\ref{fig:rho} and in the rest of the paper we choose the normalization for which $a_0\equiv a(T_0)= 1$.

This non-standard scenario tends to converge to the usual radiation dominated case when $\xi$ takes small values.
In fact, if $\xi\ll\ximin$, where
\begin{equation}\label{eq:minxsi}
\ximin\approx\left[\left(\frac{g_\star(\Tmax)}{g_\star(\Tdec)}\right)^\frac14\frac{\Tmax}{\Tdec}\right]^{3\op-1},
\end{equation}
the period when the SM energy density scales like $a^{-\frac32(1+\op)}$ tends to disappear.%
\footnote{In appendix~\ref{sec:appendix} the criterion for defining $\ximin$ is presented.}
If $\op<1/3$, this corresponds to the case where $\rho_R$ is always subdominant with respect to $\rp$.
In the opposite case, when $\op>1/3$, $\phi$ decays when its energy density is already subdominant.

\section{Primordial Gravitational Waves in Non-standard Cosmologies}
\label{sec:pgw-non}

In this paper we consider scenarios where for some period at early times the expansion 
of the Universe was governed by a fluid component with an effective equation of state $\op$.
Particular cases correspond to $\op=-1$ (quintessence), $0$ (matter, modulus), $1/3$ (radiation), 1 (kination); however we consider general cases where $\op\in[0,\,1]$ in our numerical analysis.
During the epoch when $\phi$ dominates the energy density of the Universe, the scale factor goes like $a(u)\propto u^{\frac{2}{3\op+1}}$, in contrast to the standard case (i.e. radiation dominated), where $a(u)\propto u$.
Therefore, the friction term in eq.~\eqref{eq:gwode} leads to more or less damping than in the usual radiation case.
It can be estimated to be
\begin{equation}
    \frac{2}{a(u)}\frac{da}{du}\sim \frac{4}{3\op+1}\frac{1}{u}\,,
\end{equation}
so that for $\op>1/3$ the friction term is reduced with respect to the usual scenario. 
In these cases, the spectrum of GW can be enhanced.
In the next sections we also consider the effect of tensor tilt $n_T$ on the power spectrum which can boost or damp the power spectrum at high frequencies. 

We should emphasize that these non-standard cosmological scenarios are viable from the perspective of CMB data.
This can be identified by using the range of variation of the number of $e$-folds $N$ on the scalar spectral index $n_s$ and the tensor-to-scalar ratio $r$.
For slow-roll inflation, the current limit on $N$ from Planck is $50\lesssim N\lesssim 60$~\cite{Akrami:2018odb}.
The equation of state parameter $\op$ for the fluid $\phi$ dominated after inflation should satisfy $|\Delta N|\lesssim10$ using the uncertainties from CMB data.%
\footnote{The uncertainties in $n_s$ and $r$ related to the scale dependency of $n_s$ can be written as~\cite{Easther:2013nga,Kinney:2005in,Adshead:2010mc}
\begin{eqnarray}
	\Delta n_s&\approx& (n_s -1)\left[-\frac{5}{16}r-\frac{3}{64}\frac{r^2}{n_s-1}\right]\Delta N,\\
	\Delta r &\approx& r\left[n_s-1+\frac{r}{8}\right]\Delta N.
\end{eqnarray}
Using Planck data~\cite{Akrami:2018odb}, previous equations impose a limit on $\Delta N$ given by $|\Delta N|\lesssim10$.
}
The change in the number of $e$-folds due to non-standard scenarios after inflation is~\cite{Akrami:2018odb, Easther:2013nga, Kinney:2005in, Adshead:2010mc}
\begin{equation}\label{eq:DN1}
	\Delta N\approx \frac{1-3\op}{12(1+\op)}\ln \frac{\rR(\Tend)}{\rho_{\text{rh}}},
\end{equation}
where $\rho_{\text{rh}}$ is the total energy density after reheating. 
If the Universe is dominated by a scalar field (which could also be the inflaton itself), one has $\rho_{\text{rh}} =M_{\phi}^2\,\Delta \phi^2/2$, where $M_\phi$ corresponds to the mass of $\phi$ and $\Delta \phi \sim M_{Pl}$~\cite{Easther:2013nga,Kinney:2005in,Adshead:2010mc}.
Using eq.~\eqref{eq:Tend}, eq.~\eqref{eq:DN1} can be rewritten as
\begin{equation}
\Delta N\approx \frac{1-3\op}{12(1+\op)} \left[-125.45+ \ln 
\left[\left(\frac{\gs(\Tend)}{10.75}\right)\left(\frac{\Tend}{10~\text{MeV}}\right)^4\left(\frac{190~\text{TeV}}{M_{\phi}}\right)^2\left(\frac{M_{Pl}}{\Delta \phi}\right)^2\right]
\right],
\end{equation}
where $\Gamma_{\phi}=M_{\phi}^3/(8\pi\,M_{Pl}^2)$ was assumed.
Typical values used in this work for $\op$, $\Tend$ and $\Gamma_\phi$ agree with the CMB bound $|\Delta N|\lesssim10$.

Moreover, non-standard cosmological scenarios assuming different 
equations of state for $\phi$ can affect the growth of primordial density
 perturbations~\cite{Redmond:2018xty, Fan:2014zua, Redmond:2017tja}. 
Primordial density perturbations for subhorizon modes ($k\tau \gg1$)  grows 
like $\delta\rho/\rho \propto a^{(3\op-1)/2}$ for $\op\neq 0$ and for 
$\op=0$ as $\delta\rho/\rho \propto a$~\cite{Artymowski:2016tme,Redmond:2018xty, Fan:2014zua, Redmond:2017tja}. 
When the growth of perturbations is large, e.g. when it scales like $a$, it may boost the formation of large structures. However, 
these modes formed at temperatures higher than $1$~MeV are much smaller than the size of horizon 
 at the time of structure formation which happens at temperature around $1$~eV. 
 Due to the diffusion (Silk) damping these modes are subdominant during the 
 formation of structures and are not effective~\cite{Artymowski:2016tme,Silk:1967kq}. 
 As a consequence, the non-standard cosmologies we consider are by construction in agreement 
 with the prediction of standard cosmology after BBN.
Here we will generally focus on the non-standard cosmological scenarios and their 
impacts on the PGW spectrum and their possible bounds from GW experiments.

The non-standard cosmology can let an imprint for frequencies higher than
\begin{equation}\label{eq:fend}
\fend\equiv\frac{\kend}{2\pi}=\frac{\aend\,H_\text{end}}{2\pi}\approx\frac16\sqrt{\frac{\gs(\Tend)}{10}}\left[\frac{\hs(T_0)}{\hs(\Tend)}\right]^{1/3}\frac{a_0\,T_0\,\Tend}{M_{Pl}},
\end{equation}
where $H_\text{end}\equiv H(T=\Tend)$.
Eq.~(\ref{eq:fend}) is derived under assuming the entropy conservation for scale factors $a\gg\aend$ until today. 
On the contrary, $f\ll\fend$ corresponds to frequencies that cross the horizon after the end of the $\phi$ domination and therefore are not sensitive to the non-standard phase.
Similarly, the frequency $\feq$ corresponding to $a=\aeq$ can be defined as
\begin{equation}\label{eq:feq}
    \feq\equiv\frac{\keq}{2\pi}\approx\frac16\sqrt{\frac{\gs(\Teq)}{5}}\,\left[\frac{\hs(T_0)}{\hs(\Tdec)}\right]^\frac13\,\left[\xi^\frac{4}{3\op-1}\left(\left[\frac{\gs(\Tmax)}{\gs(\Tdec)}\right]^\frac14\frac{\Tmax}{\Tdec}\right)^{3\op-1}\right]^\frac{1}{3(1+\op)}\,\frac{a_0\,T_0\,\Teq^2}{M_{Pl}\,\Tmax},
\end{equation}
see appendix~\ref{sec:appendix} for details.

The present relic of gravitational waves can be approximately written as
\begin{equation}\label{eq:OGWapprox}
    \OGW(\eta_0,\,k)\approx\frac{\mathscr{P}_T(k)\,k^2\,a^2_\text{hc}}{24\,a^4_0\,H^2_0}\,,
\end{equation}
where $\ahc$ is the scale factor at horizon crossing, and $\eta_0$ is the conformal time today.
Considering a Universe dominated by a $\phi$ component before BBN leads to different regimes for the PGW spectrum depending on the moment where perturbations cross the horizon.
We classify them in the following.

\subsection{Classification}\label{sec:classification}
\subsubsection[Case 1: $\aend\ll\ahc$]{Case 1: $\boldsymbol{\aend\ll\ahc}$}\label{sec:1}
In this case perturbations cross the horizon well after the decay of $\phi$, when $\aend\ll\ahc$.
This corresponds to the standard scenario where the Universe is radiation dominated, and therefore the Hubble expansion rate scales like
\begin{equation}
    H(a)=\sqrt{\frac{\rR(a)}{3M_{Pl}^2}}=\tHmax\,\left(\frac{\astart}{\aend}\right)^\frac{3\op-5}{4}\,\left(\frac{\amax}{a}\right)^2,
\end{equation}
where
\begin{equation}\label{eq:Hmaxphi}
    \tHmax\equiv\frac{\pi}{3}\sqrt{\frac{\gs(\Tmax)}{10}}\,\frac{\Tmax^2}{M_{Pl}}
\end{equation}
corresponds to the contribution to $H$ coming from the SM radiation at $T=\Tmax$.
Let us emphasize that $\tHmax\neq H(T=\Tmax)$.
Additionally, the scale factor $\amax$ at $\Tmax$ can be estimated to be
\begin{equation}\label{eq:amax}
    \amax \approx a_0\,\frac{T_0}{\Tmax}\,\left[\frac{\hs(T_0)}{\hs(\Tdec)}\right]^\frac13\,\left[\xi\left(\left[\frac{\gs(\Tmax)}{\gs(\Tdec)}\right]^\frac14\frac{\Tmax}{\Tdec}\right)^{1-3\op}\right]^{-\frac{1}{3(1+\op)}},
\end{equation}
see appendix~\ref{sec:appendix}.
Therefore, at the horizon crossing
\begin{equation}
    k=\ahc\,H(\ahc)=\tHmax\,\left(\frac{\astart}{\aend}\right)^\frac{3\op-5}{4}\,\frac{\amax^2}{\ahc}\propto\ahc^{-1}\,.
\end{equation}
That dependence implies that $\OGW$ will inherit exactly the same scale dependence as the primordial spectrum
\begin{equation}
    \OGW(\eta_0,\,k)\approx\frac{\mathscr{P}_T(k)}{24}\left(\frac{\tHmax}{H_0}\right)^2\left(\frac{\amax}{a_0}\right)^4\left(\frac{\astart}{\aend}\right)^\frac{3\op-5}{2}\propto\mathscr{P}_T(k).
\end{equation}
In particular, if the primordial spectrum is scale invariant, $\OGW$ becomes independent of $k$, up to changes in the relativistic degrees of freedom.

\subsubsection[Case 2: $\aeq\ll\ahc\ll\aend$]{Case 2: $\boldsymbol{\aeq\ll\ahc\ll\aend}$}\label{sec:2}
This case corresponds to the scenario where $\aeq\ll\ahc\ll\aend$. Additionally, we demand that $\xi\gg\ximin$, which implies that a sizable relative increase of the temperature is achieved due to the decay of $\phi$.
This is typically realized when $\op\ll 1/3$ and therefore $\Tend\approx\Tdec$.
In this case $\phi$ dominates the Hubble expansion rate\footnote{Let us note that if $\op>1/3$, for $a<\aend$ the Universe is always dominated by $\phi$.}, and therefore
\begin{eqnarray}\label{eq:hubblephi}
    H(a)&=&\sqrt{\frac{\rp(a)}{3M_{Pl}^2}}=\sqrt{\frac{\rp(\amax)}{3M_{Pl}^2}}\left(\frac{\amax}{a}\right)^{\frac32(1+\op)}\nonumber\\
    &=&\frac{\pi}{3}\sqrt{\frac{\gs(\Tmax)}{10}}\,\frac{\Tmax^2}{M_{Pl}}\,\sqrt{\xi}\,\left(\frac{\amax}{a}\right)^{\frac32(1+\op)}=\tHmax\,\sqrt{\xi}\,\left(\frac{\amax}{a}\right)^{\frac32(1+\op)}.
\end{eqnarray}
This implies that at the horizon crossing
\begin{equation}\label{eq:kphi}
    k=\ahc\,H(\ahc)=\tHmax\,\sqrt{\xi}\,\amax^{\frac32(1+\op)}\,\ahc^{-\frac{1+3\op}{2}}.
\end{equation}
That allows to find an approximate expression for the present relic of GW:
\begin{equation}\label{eq:relicphi}
    \OGW(\eta_0,\,k)\approx\frac{\mathscr{P}_T(k)}{24\,a^4_0\,H^2_0}\,\left[\tHmax^2\,\xi\,\amax^{3(1+\op)}\,k^{3\op-1}\right]^\frac{2}{1+3\op},
\end{equation}
which presents an extra $k$-dependence, additionally to the one from the primordial spectrum.

\subsubsection[Case 3: $\ahc\ll\aeq$]{Case 3: $\boldsymbol{\ahc\ll\aeq}$}\label{sec:3}
This case corresponds to the scenario where $\ahc\ll\aeq$.
We again demand that $\xi\gg\ximin$, which implies that a sizable relative increase of the temperature is achieved due to the decay of $\phi$.
This can only be realized when $\op<1/3$ and therefore $\Tend\approx\Tdec$.
In this case the Universe is radiation dominated and then the  Hubble rate evolves like
\begin{eqnarray}\label{eq:hubbleomega0}
    H(a)&=&\sqrt{\frac{\rho_R(a)}{3M_{Pl}^2}}=\sqrt{\frac{\rho_R(\amax)}{3M_{Pl}^2}}\left(\frac{\amax}{a}\right)^{2}\nonumber\\
    &=&\frac{\pi}{3}\sqrt{\frac{\gs(\Tmax)}{10}}\,\frac{\Tmax^2}{M_{Pl}}\left(\frac{\amax}{a}\right)^{2}=\tHmax\left(\frac{\amax}{a}\right)^{2}\,.
\end{eqnarray}
Here the horizon crossing happens for
\begin{equation}\label{eq:hcomega0}
    k=\ahc\,H(\ahc)=\tHmax\,\frac{\amax^2}{\ahc}.
\end{equation}
Then the relic of PGW for radiation domination can be estimated to be
\begin{equation}\label{eq:omega0}
    \OGW(\eta_0,\,k)\approx\frac{\mathscr{P}_T(k)}{24}\,\left(\frac{\tHmax}{H_0}\right)^2\,\left(\frac{\amax}{a_0}\right)^4,
\end{equation}
where $\amax$, given in eq.~\eqref{eq:amax}, is the only place where a $\xi$ dependence appears. 
As expected, eq.~\eqref{eq:omega0} only depends on $k$ via the primordial spectrum $\mathscr{P}_T(k)$.

\subsubsection[Case 4:  $\xi\ll\ximin$]{Case 4:  $\boldsymbol{\xi\ll\ximin}$}\label{sec:4}
This last case corresponds to the scenario where $\xi\ll\ximin$, which implies that either $\rp$ is subdominant when $\phi$ decays, or that $\phi$ is not decaying at all.
This is typically realized when $\op\gg 1/3$.\footnote{In fact, for $\op<1/3$ and $\xi\ll\xi_{\text{min}}$ the Universe is always radiation dominated, and hence corresponds to the standard cosmology.}
Additionally, here $\Tend\gg\Tdec$.
Let us also note that in this case $\aeq$ and $\astart$ are not defined, so the only relevant scale is $a=\aend$.
The scenario where $\ahc\gg\aend$ corresponds to the previously discussed case~\hyperref[sec:1]{1}, now we focus on the opposite case $\ahc\ll\aend$.

In this scenario, as the energy density is dominated by $\rp$, the Hubble expansion rate is given by eq.~\eqref{eq:hubblephi}.
However, the scale factor $\amax$ can now be computed by the use of the conservation of the SM entropy, which implies that
\begin{equation}\label{eq:amax13}
    \amax\approx a_0\,\frac{T_0}{\Tmax}\,\left[\frac{\hs(T_0)}{\hs(\Tmax)}\right]^\frac13,
\end{equation}
which is now independent of $\xi$.
Similar to case~\hyperref[sec:1]{1}, the horizon crossing and the present relic of GW are given by eqs.~\eqref{eq:kphi} and~\eqref{eq:relicphi}, respectively.
Additionally, using eq.~\eqref{eq:fend} and entropy conservation, in this case\footnote{Note that in this scenario $\Tend\approx\Tmax\left[\frac{\hs(\Tmax)}{\hs(\Tend)}\right]^{1/3}\xi^{\frac{1}{3\op-1}}$.}
\begin{equation}\label{eq:fend2}
    \fend=\frac{\aend\,H_\text{end}}{2\pi}\approx\frac16\,\left[\frac{\hs(T_0)\,\hs(\Tmax)}{\hs(\Tend)^2}\right]^\frac13\sqrt{\frac{\gs(\Tend)}{10}}\,\xi^\frac{1}{3\op-1}\,\frac{a_0\,T_0\,\Tmax}{M_{Pl}}.
\end{equation}


\subsection{Scale Invariant Primordial Spectrum}
\label{sec:exp-scale}
\begin{figure}[t]
\begin{center}
\includegraphics[height=0.46\textwidth]{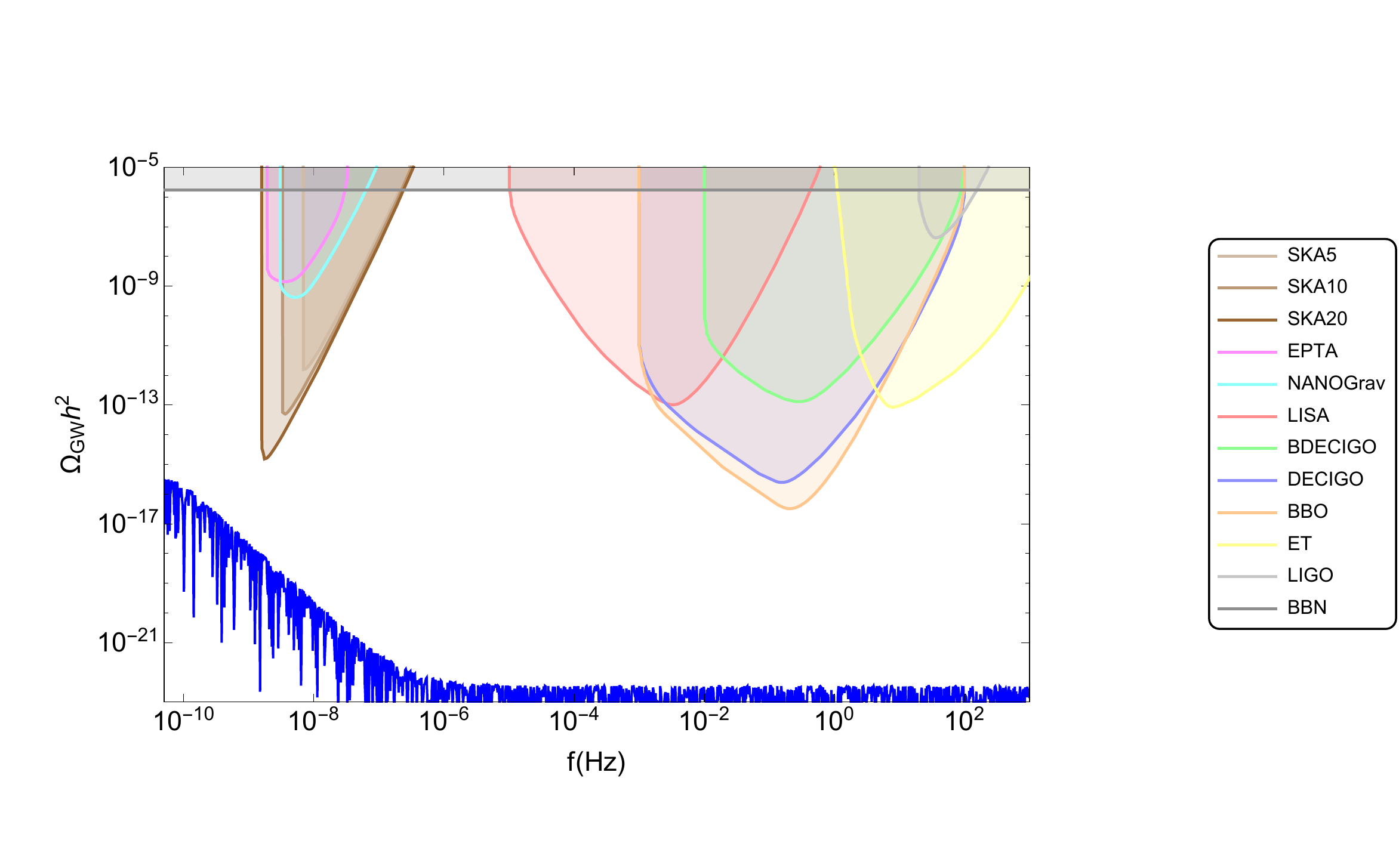}\\
\vspace{-1.5cm}
\includegraphics[height=0.46\textwidth]{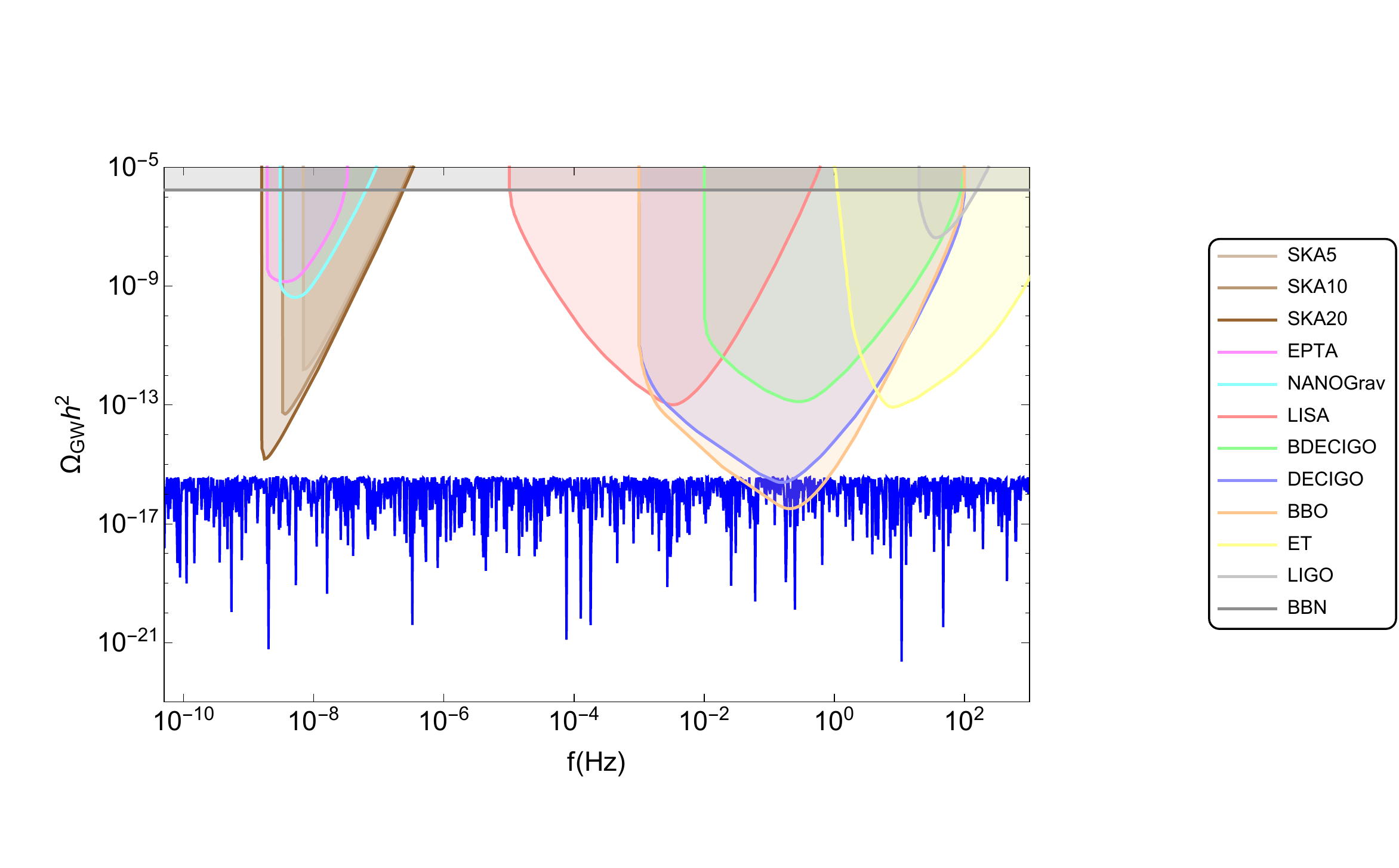}\\
\vspace{-1.5cm}
\includegraphics[height=0.46\textwidth]{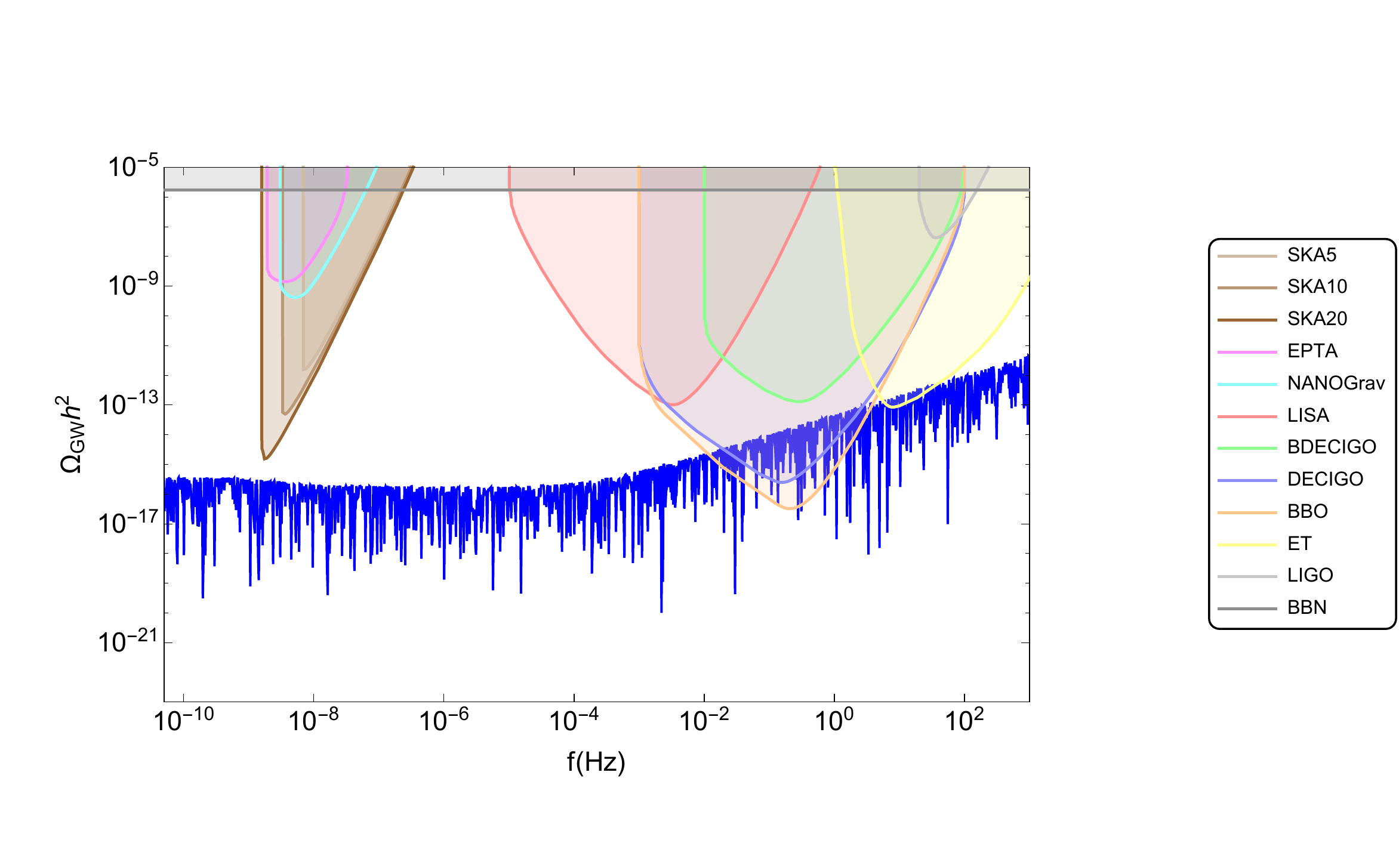}
\caption{Spectra of inflationary GWs for $\op=0$ and $\xi=10^{-11}$ (upper panel), $\op=1/3$ and $\xi=10^{25}$ (central panel), and $\op=2/3$ and $\xi=10^{10}$ (lower panel).
$\Tdec=10$~MeV was also chosen.
These benchmark points are the same used in fig.~\ref{fig:rho} and are shown in the upper left panel of fig.~\ref{fig:treh-vinf}. 
The colored regions correspond to projected sensitivities for various gravitational wave observatories, and to the BBN constraint described in the text.
}
\label{fig:OGW}
\end{center}
\end{figure}
\begin{figure}[t]
\begin{center}
\includegraphics[height=0.46\textwidth]{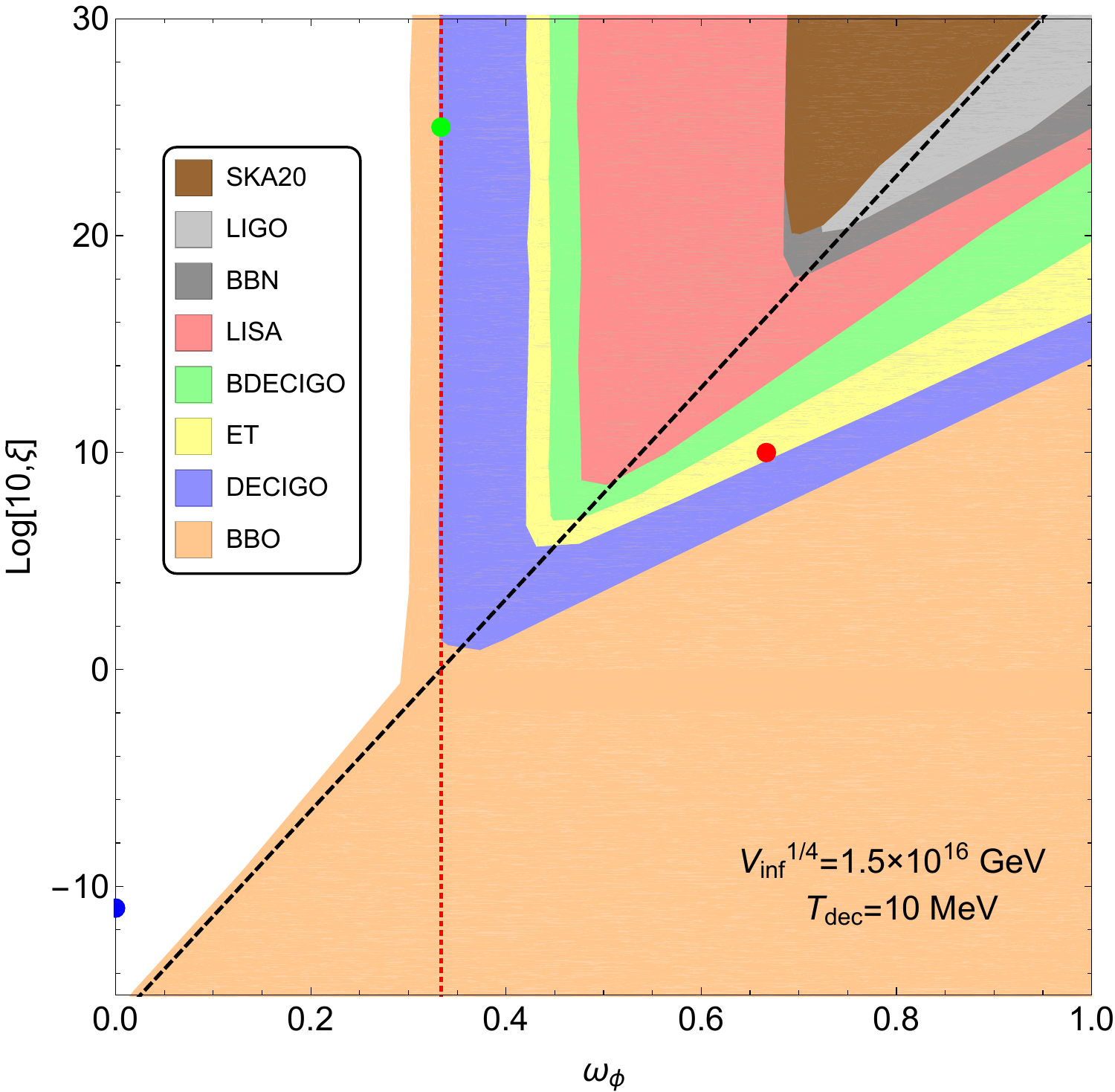}
\includegraphics[height=0.46\textwidth]{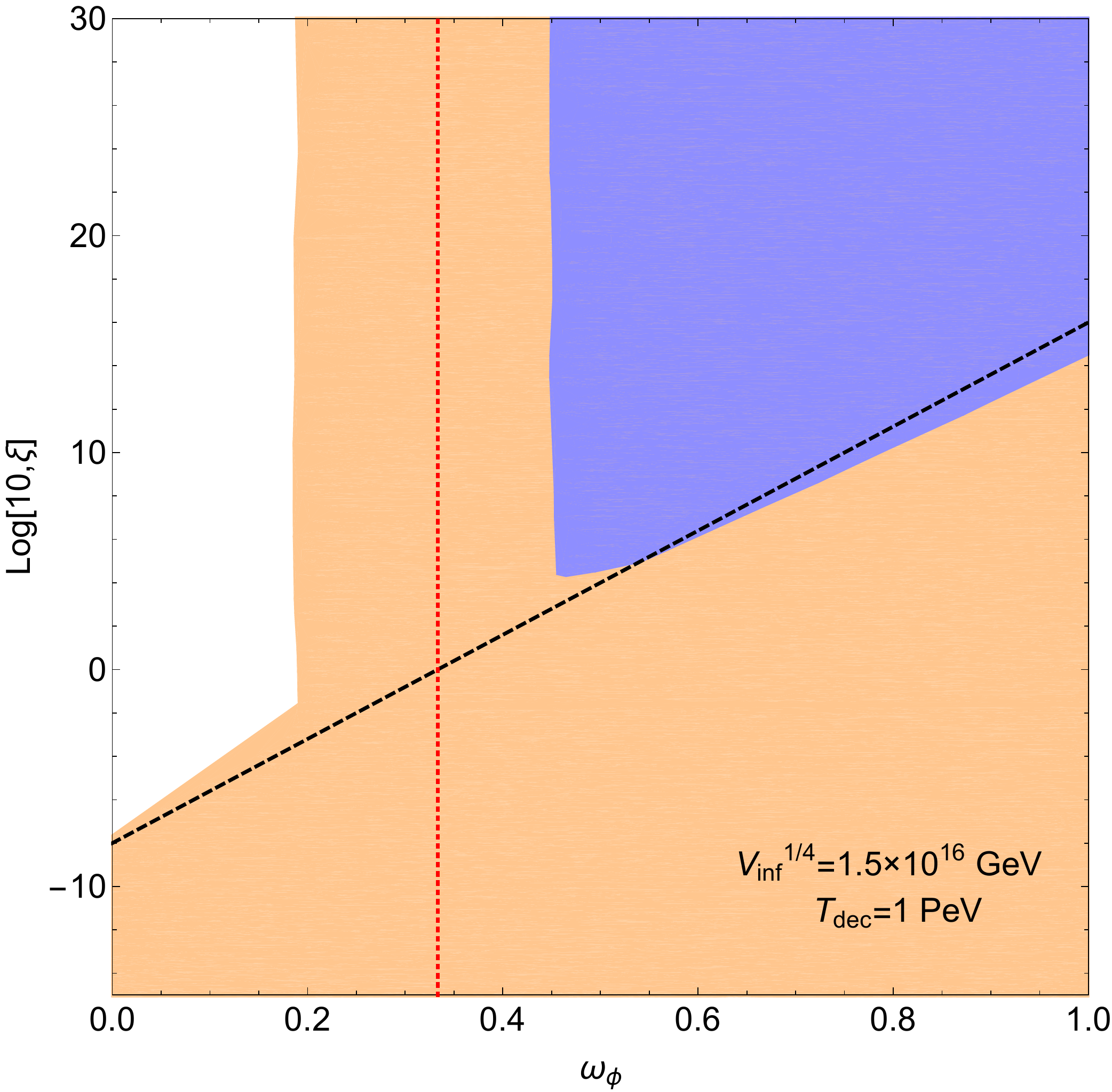}
\includegraphics[height=0.46\textwidth]{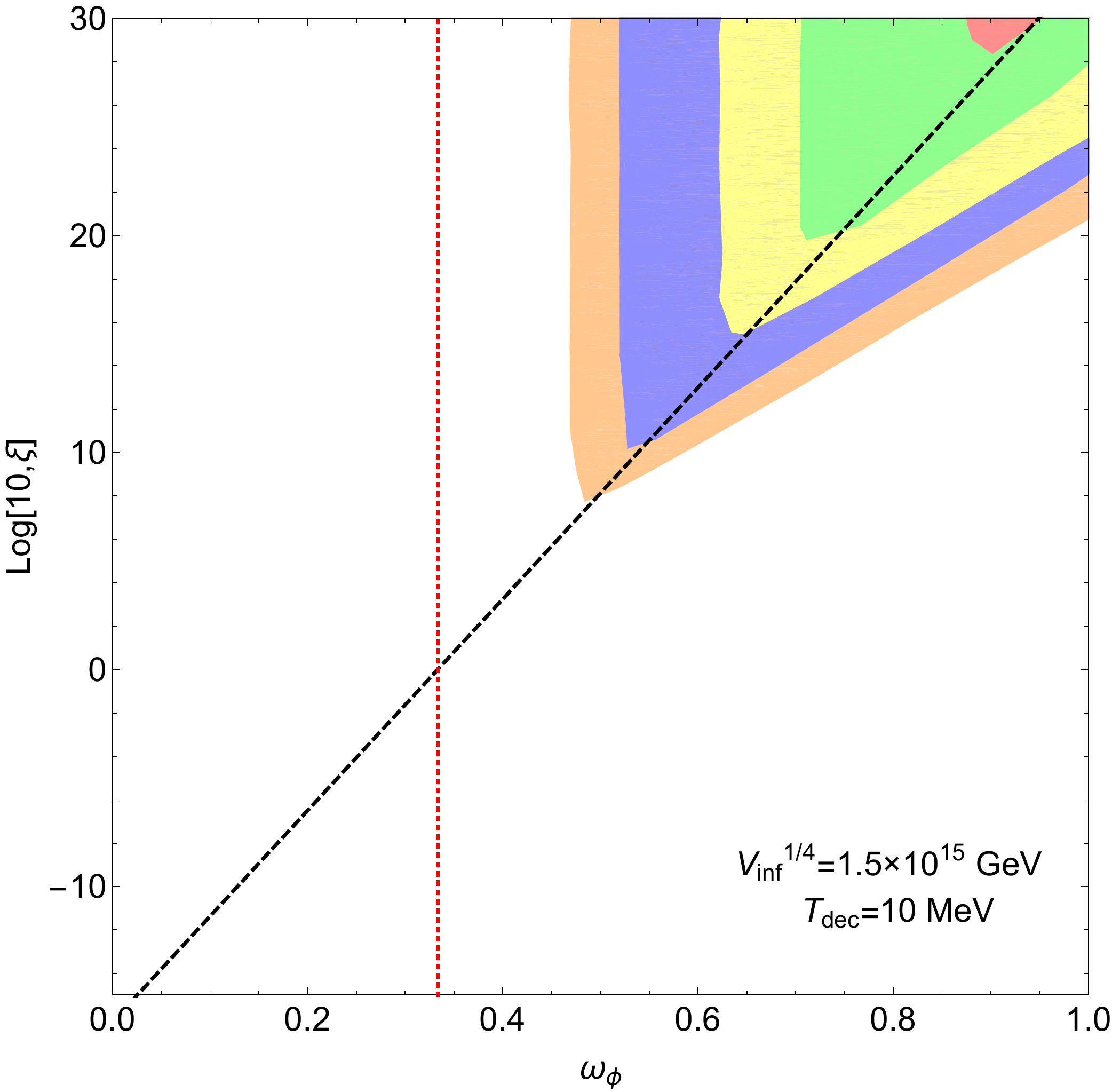}
\includegraphics[height=0.46\textwidth]{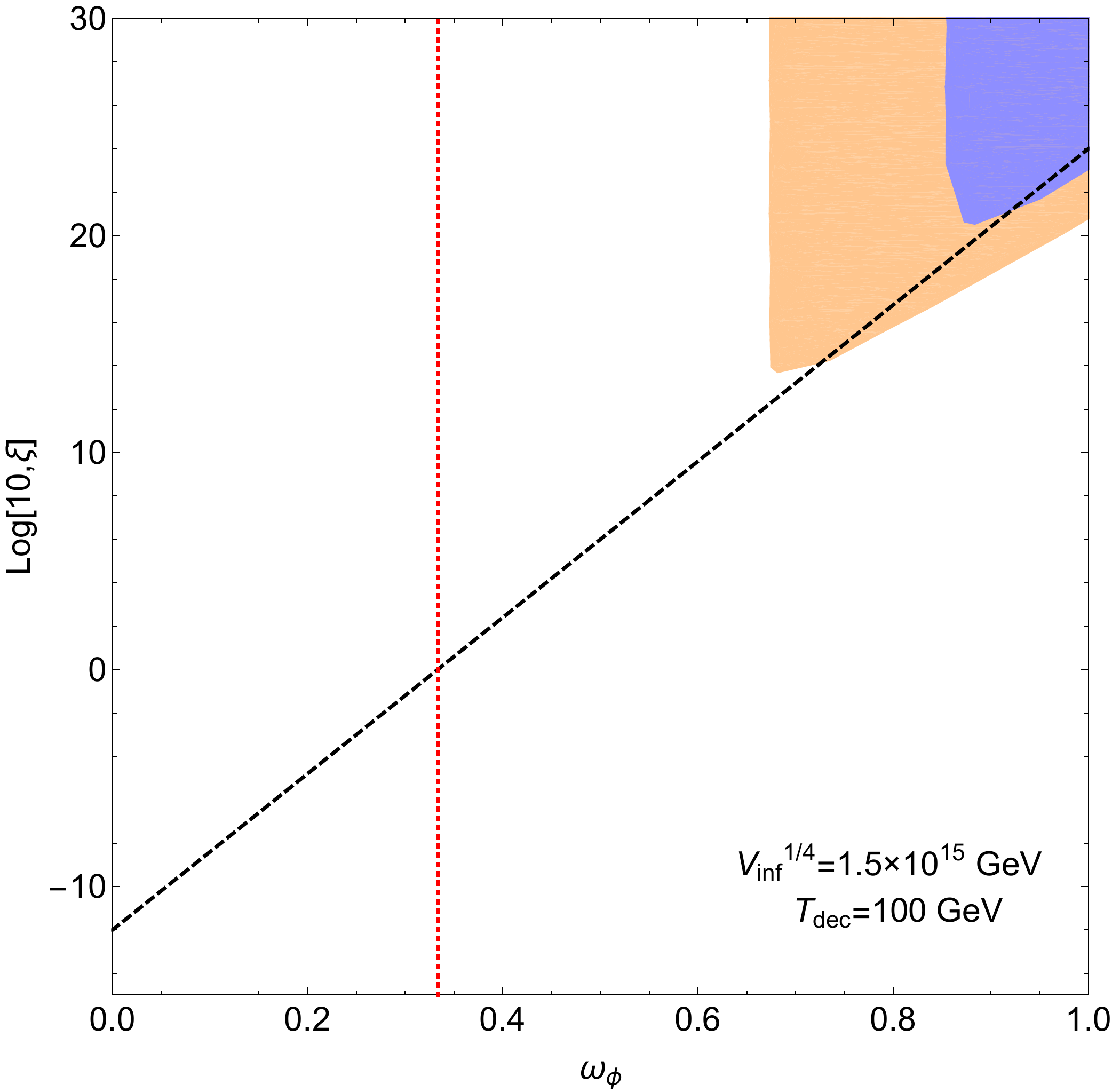}
\caption{Regions of the parameter space that could be tested by different observatories for a scale invariant primordial spectrum ($n_T=0$), taking $V_\text{inf}^{1/4}=1.5\times10^{16}$~GeV (upper panels) and $1.5\times10^{15}$~GeV (lower panels), and $\Tdec=10$~MeV (left panels),  1~PeV (upper right panel) and 100~GeV (lower right panel).
Three colored markers in the upper left panel are benchmark points for figs.~\ref{fig:rho} and~\ref{fig:OGW}.
The diagonal black dashed lines correspond to $\ximin$.
The vertical red dotted lines correspond to $\op=1/3$.
}
\label{fig:treh-vinf}
\end{center}
\end{figure}

Figure~\ref{fig:OGW} shows examples of spectra of inflationary GWs.
The upper panel corresponds to $\op=0$ and $\xi=10^{-11}$, the central panel to $\op=1/3$ and $\xi=10^{25}$ and the lower panel to $\op=2/3$ and $\xi=10^{10}$.
In all panels the temperature at which $\phi$ decays is assumed to be $\Tdec=10$~MeV.
Here we are considering a primordial scale invariant spectrum ($n_T=0$) with $V_\text{inf}^{1/4}=1.5\times 10^{16}$~GeV.
Let us note that the benchmarks are the same used in fig.~\ref{fig:rho}, and presented in the upper left panel of fig.~\ref{fig:treh-vinf}.

In the upper panel of fig.~\ref{fig:OGW} we assumed $\op=0$.
For frequencies smaller than $\fend\sim 10^{-10}$~Hz (eq.~\eqref{eq:fend}),
perturbations crossed the horizon after the end of the $\phi$ domination and therefore are not sensitive to the non-standard phase, case~\hyperref[sec:1]{1}.
The GW spectrum is therefore scale invariant, as the primordial spectrum.
For higher frequencies, $\phi$ dominates the Hubble expansion rate and therefore $\OGW\propto f^{-2\frac{1-3\op}{1+3\op}}=f^{-2}$, case~\hyperref[sec:2]{2}.
For $f>\feq\sim 10^{-6}$~Hz (eq.\eqref{eq:feq}),
the Universe is again radiation dominated and therefore the spectrum becomes again scale invariant, case~\hyperref[sec:3]{3}.\\
In the central panel of fig.~\ref{fig:OGW} we took  $\op=1/3$, implying that the Universe is always dominated by a component that scales like radiation: either the SM radiation or $\phi$.
The GW spectrum has therefore the same $k$ dependence as the primordial spectrum which is scale invariant.\\
Finally, in the lower panel of the figure $\op=2/3$ and $\xi<\ximin$.
For $f<\fend\sim 10^{-4}$~Hz (eq.~\eqref{eq:fend2}),
the GW spectrum is essentially flat, up to variations due to the change of the relativistic degrees of freedom.
For $f>\fend$, the GW spectrum is modified by the non-standard phase and scales like $\OGW\propto f^{-2\frac{1-3\op}{1+3\op}}=f^\frac23$, case~\hyperref[sec:4]{4}.

Additionally, in fig.~\ref{fig:OGW} the colored regions correspond to projected sensitivities for various gravitational wave observatories~\cite{Breitbach:2018ddu}.
In particular, we consider the proposed ground-based Einstein Telescope (ET)~\cite{Sathyaprakash:2012jk}, the planned space-based LISA~\cite{Audley:2017drz} interferometer as well as the proposed successor experiments BBO~\cite{Crowder:2005nr} and (B-)DECIGO~\cite{Seto:2001qf, Sato:2017dkf}.
Moreover, we include pulsar timing arrays, in particular the currently operating EPTA~\cite{Lentati:2015qwp} and NANOGrav~\cite{Arzoumanian:2018saf}, as well as the future SKA~\cite{Janssen:2014dka} telescope. 
For the frequency range $10^{-3}$ to $10^2$~Hz the $\OGW h^2\sim2\times10^{-17}$ is the lowest relic that can be probed by BBO experiment.
DECIGO can probe GW relics above $\sim2\times10^{-16}$ in the same frequency range.
Moreover, BDECIGO can probe frequencies between $10^{-2}$ and $10^2$~Hz with a maximum sensitivity around $10^{-13}$.
A similar sensitivity could also be reached by LISA but in the frequency range
$10^{-5}$ to $1$~Hz. Very large frequencies between $1$ and $10^4$~Hz will be probed by ET experiment for $\OGW h^2\sim2\times10^{-13}$. NANOGrav and EPTA can probe 
regions between $10^{-9}$ and $10^{-7}$~Hz with relic above $10^{-9}$.
Finally, SKA can probe the regions between $10^{-9}$ and $10^{-6}$ Hz depending on the period of operation for 5, 10, and 20 years.
The constraints from LIGO/VIRGO collaboration on the stochastic 
gravitational background and the coalescence of compact binary objects
are also considered in our analysis~\cite{Abbott:2017xzg, LIGOScientific:2019vic}.
A primordial gravitational wave relic as small as $\sim 3\times 10^{-8}$ is not observed and therefore excluded for the frequency range between $10$ and $200$~Hz.
Additionally, the PGW background as an extra radiation component modifies the expansion rate of Universe and can therefore be constrained by BBN~\cite{Boyle:2007zx}.
This is done by using the measurement of the number of effective neutrinos $N_\text{eff}$ and the observational abundance of $D$ and $^4He$, which impose $\OGW h^2<\Omega_\text{BBN}h^2\simeq 1.7\times10^{-6}$ at $95\%$~CL~\cite{Stewart:2007fu,Kohri:2018awv} which shows the integrated amount of PGW radiation.
Combining the constraint from BBN on PGW background and eq.~\eqref{eq:relicphi} we can find the maximum frequency at which the Universe can start to be $\phi$ dominated:
\begin{equation}
    k_\text{BBN}\approx\left[\left(36\pi^2\frac{a^4_0\,M_{Pl}^4\,H^2_0\,\Omega_\text{BBN}}{V_{\text{inf}}}\right)^{\frac{3\op+1}{2}}\tHmax^{-2}\,\xi^{-1}\,\amax^{-3(1+\op)}\right]^{\frac{1}{3\op-1}}.
\end{equation}
For frequencies larger than $f_{\text{BBN}}$ the Universe should be either radiation dominated or during an inflationary phase.
Other indirect possible constraints on PGW include the effect on temperature and polarization of the CMB, and matter power spectra considered in refs.~\cite{Pagano:2015hma, Lasky:2015lej}.
However, these limits are not as competitive as the BBN bound.

Figure~\ref{fig:treh-vinf} shows in colors the regions of the parameter space $[\op,\,\xi]$ that could be probed by different observatories for a scale invariant primordial spectrum ($n_T=0$), taking $V_\text{inf}^{1/4}=1.5\times10^{16}$~GeV (upper panels) and $1.5\times10^{15}$~GeV (lower panels), and $\Tdec=10$~MeV (left panels), 1~PeV (upper right panel) and 100~GeV (lower right panel). 
The lines correspond to $\xi=\ximin$ (black dashed lines) and $\op=1/3$ (red dotted vertical lines).
Some general comments are in order.
There is an important lost of sensitivity when decreasing the scale of inflation $V_\text{inf}^{1/4}$, because the GW spectrum $\OGW\propto \mathscr{P}_T\propto V_{\text{inf}}$.
However, sensitivity increases when maximizing the $\phi$-dominance period by decreasing $\Tdec$.
One can see that different experiments could probe complementary regions of the parameter space, typically corresponding to equations of state $\op>1/3$ and to $\xi>\ximin$.

The behavior of the sensitivity regions can be understood as follows.
Typically, the minimum value of equation of state parameter $\op$ that can be probed by a given experiment happens in case~\hyperref[sec:2]{2}, when $\aeq\ll\ahc\ll\aend$ and $\xi>\ximin$.
It can be found by equaling $\OGW$ (eq.~\eqref{eq:relicphi}) with $\Omega_\text{min}$, so that
\begin{equation}\label{appcons1}
    \omega_{\phi,\,\text{min}}\approx \frac43\frac{\ln \left[\left(\frac{\hs(T_0)}{\hs(\Tdec)}\right)^\frac13\left(\frac{\gs(\Tdec)}{\gs(\Tmax)}\right)^\frac14\frac{a_0\,\tHmax\,\Tdec\,T_0}{\Tmax^2}\right]}{\ln \left[\left(\frac{\hs(\Tdec)}{\hs(T_0)}\right)^\frac23\left(\frac{\gs(\Tdec)}{\gs(\Tmax)}\right)^\frac12\frac{36 \pi^2\,a_0^2\,M_{Pl}^4\,H_0^2\,\Omega_{\text{min}}\,\Tdec^2}{V_\text{inf}\, k_\text{min}^2\,T_0^2}\right]}-\frac13\,,
\end{equation}
where $\Omega_\text{min}\equiv \OGW(k_\text{min})$ corresponds to the maximal sensitivity that a given experiment can reach, and $k_\text{min}$ to the wave number at which $\Omega_\text{min}$ occurs.
The fact that eq.~\eqref{appcons1} is independent of $\xi$ explains that the bounds on fig.~\ref{fig:treh-vinf} appear as vertical lines.

However, the parameter space corresponding to $\op<\omega_{\phi,\,\text{min}}$ could also be probed.
This corresponds to case~\hyperref[sec:3]{3}, where $\ahc\ll\aeq$ and $\xi>\ximin$.
The maximum $\xi$ that can be probed by a given experiments can be derived from eq.~\eqref{eq:omega0} to be
\begin{equation}\label{eq:ximaxomegaphi}
 \xi\approx\left[\frac{\gs(\Tdec)\Tdec^4}{\gs(\Tmax)\Tmax^4}\right]\left[\left(\frac{\hs(T_0)}{\hs(\Tdec)}\right)^\frac43\frac{V_\text{inf}}{36\pi^2\,\Omega_{\text{min}}\,M_{Pl}^4}\left(\frac{\tHmax}{H_0}\right)^2\frac{T_0^4}{\Tdec^4}\frac{\gs(\Tmax)}{\gs(\Tdec)}\right]^{\frac{3(1+\op)}{4}}
\end{equation}
and can be seen in the BBO bound for $\op\lesssim1/3$ in the two upper panels of fig.~\ref{fig:treh-vinf}.

Finally, the lower limit on $\xi$ that can be explored with a GW observatory typically corresponds to case~\hyperref[sec:4]{4}, where $\xi<\ximin$ and $\op>\omega_{\phi,\,\text{min}}$.
In this scenario
\begin{equation}\label{eq:xiomegaphi}
\xi\approx\left[\frac{k_{\text{min}}}{\tHmax\,\amax}\right]^2\left[\frac{36 \pi^2\,a_0^4\,H_0^2\,M_{Pl}^4\,\Omega_{\text{min}}}{V_\text{inf}\,k_\text{min}^2\,\amax^2}\right]^{\frac{1+3\op}{2}},
\end{equation}
which corresponds to the tilted colored lines for $\op>1/3$ in fig.~\ref{fig:treh-vinf} and represents the typical minimum value of $\xi$ that a given experiment can probe.

\subsection{Effect of the Tensor Tilt}\label{sec:tentilt}

In the previous section the effect of non-standard cosmologies on scale invariant primordial spectra was studied.
Here we generalize that analysis to spectra with non-zero tensor tilts~\cite{Abbott:2009ws}. 
The case of a primordial tensor power spectrum which is not scale invariant, having a $k$-dependence is usually parameterized in the following manner~\cite{Aghanim:2018eyx}:
\begin{equation}\label{ptat}
    \mathscr{P}_T(k)=A_T \left(\frac{k}{\tilde{k}}\right)^{n_T},
\end{equation}
where $A_T=\frac{2}{3\pi^2}\frac{V_\text{inf}}{M_{Pl}^4}$ is the tensor amplitude at some pivot scale $\tilde k$ and $n_T$ is the tensor spectral index.
In general, $A_T$ and $n_T$ are independent parameters.
However, in the single-field slow-roll scenario an interesting consistency relation holds between these quantities.
The tensor-to-scalar ratio~\cite{Guzzetti:2016mkm, Baumann:2009ds}
\begin{equation}\label{ratas}
    r\equiv\frac{A_T}{A_S}\,,
\end{equation}
yields the amplitude of the GW with respect to that of the scalar perturbations at some fixed pivot scale $\tilde k$, where $A_S\simeq 2.1\times 10^{-9}$~\cite{Aghanim:2018eyx} corresponds to the amplitude of primordial spectrum of curvature perturbations. 
At the lowest order in slow-roll parameters, the following consistency relation holds:
\begin{equation}\label{rnt}
    r=-8\,n_T.
\end{equation}

For a radiation dominated Universe before BBN and assuming the previous consistency relation, we scan over the parameter space of $n_T$ and $r$ to show the regions that could be constrained by GW experiments.
 \begin{figure}[t]
\begin{center}
\includegraphics[height=0.46\textwidth]{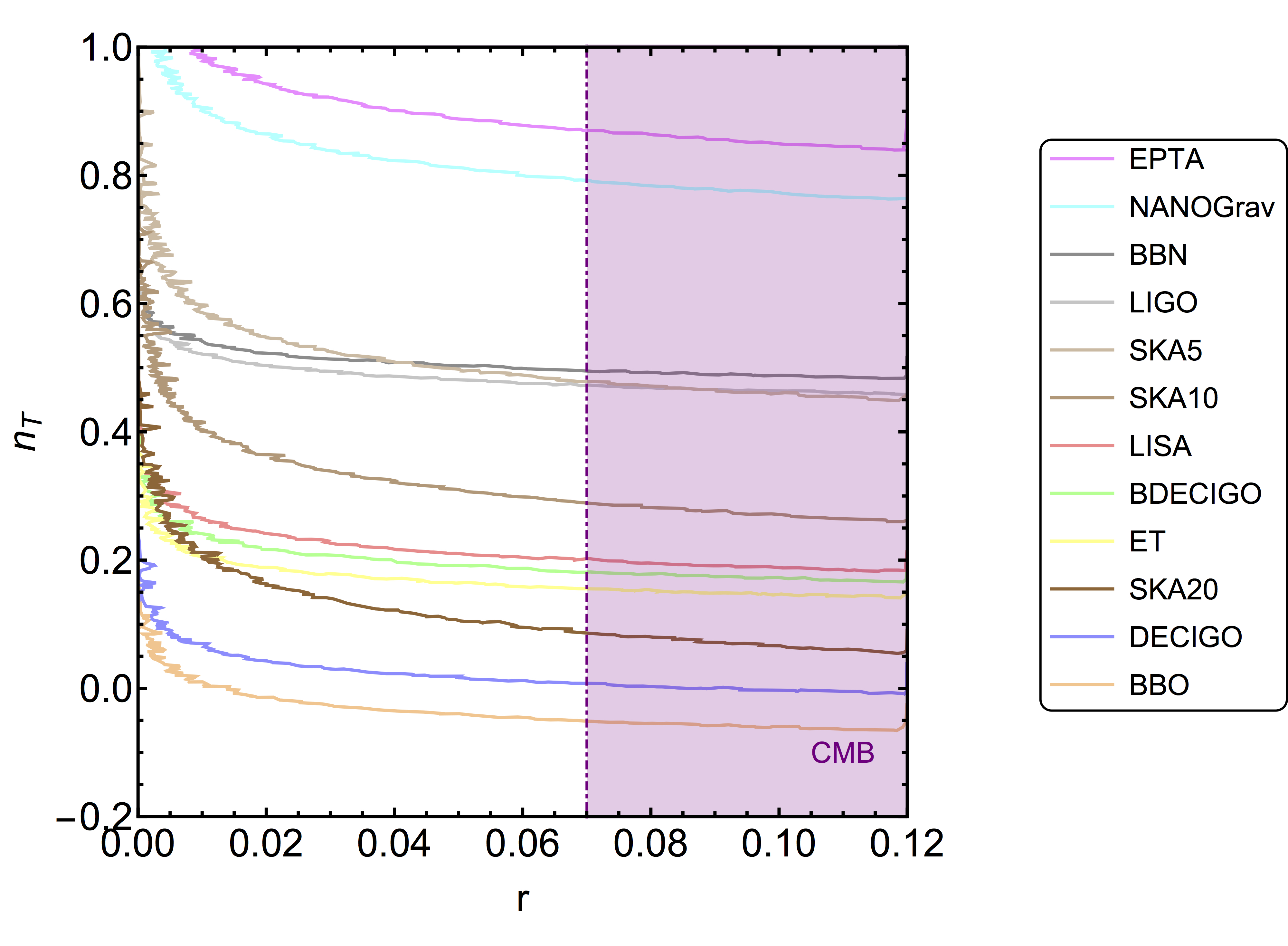}
\caption{Regions in the parameter space of the tensor-to-scalar ratio $r$ and the tensor tilt $n_T$ that can be probed by GW wave experiments in case of the standard radiation domination scenario.
The regions above the colorful curves can be potentially excluded by different experiments. There are some comments in the right of plot in order, based on the minimum value of the tensor tilt that can be probed by each experiment. 
The dot-dashed purple line shows the upper bound on $r$ from CMB by Planck satellite~\cite{Akrami:2018odb}.
}
\label{fig:r-nT}
\end{center}
\end{figure}
Figure~\ref{fig:r-nT} shows the upper bounds on the parameter space of  $[r,\,n_T]$ that can be probed by different GW experiments.
The current BBN  and LIGO bounds already constrain $n_T\gtrsim 0.5$ for $r\gtrsim 0.01$. 
The minimum value for the tensor tilt that can be probed by a given experiment can be approximated from eqs.~\eqref{gwrelic-smdof}, \eqref{ptat} and~\eqref{ratas} as
\begin{equation}\label{ntmin}
n_{T,\,\text{min}}\approx \frac{\ln \left[\frac{48}{r\,A_S\,g_\star({T_{hc}})} \frac{\Omega_{\text{min}}}{\Omega_{\gamma,\,0}} \left[\frac{\hs(\Thc)}{\hs(T_0)}\right]^{4/3}\right]}{\ln \left[\frac{k_{\text{min}}}{\tilde{k}}\right]}\,,
\end{equation}
 which has a logarithmic dependence on $r$.
PGW observatories could probe large regions on $[r,n_T]$ plane and eventually put constraints, stronger than the current CMB constraint limits~\cite{Akrami:2018odb}.

\begin{figure}[t]
\begin{center}
\includegraphics[height=0.46\textwidth]{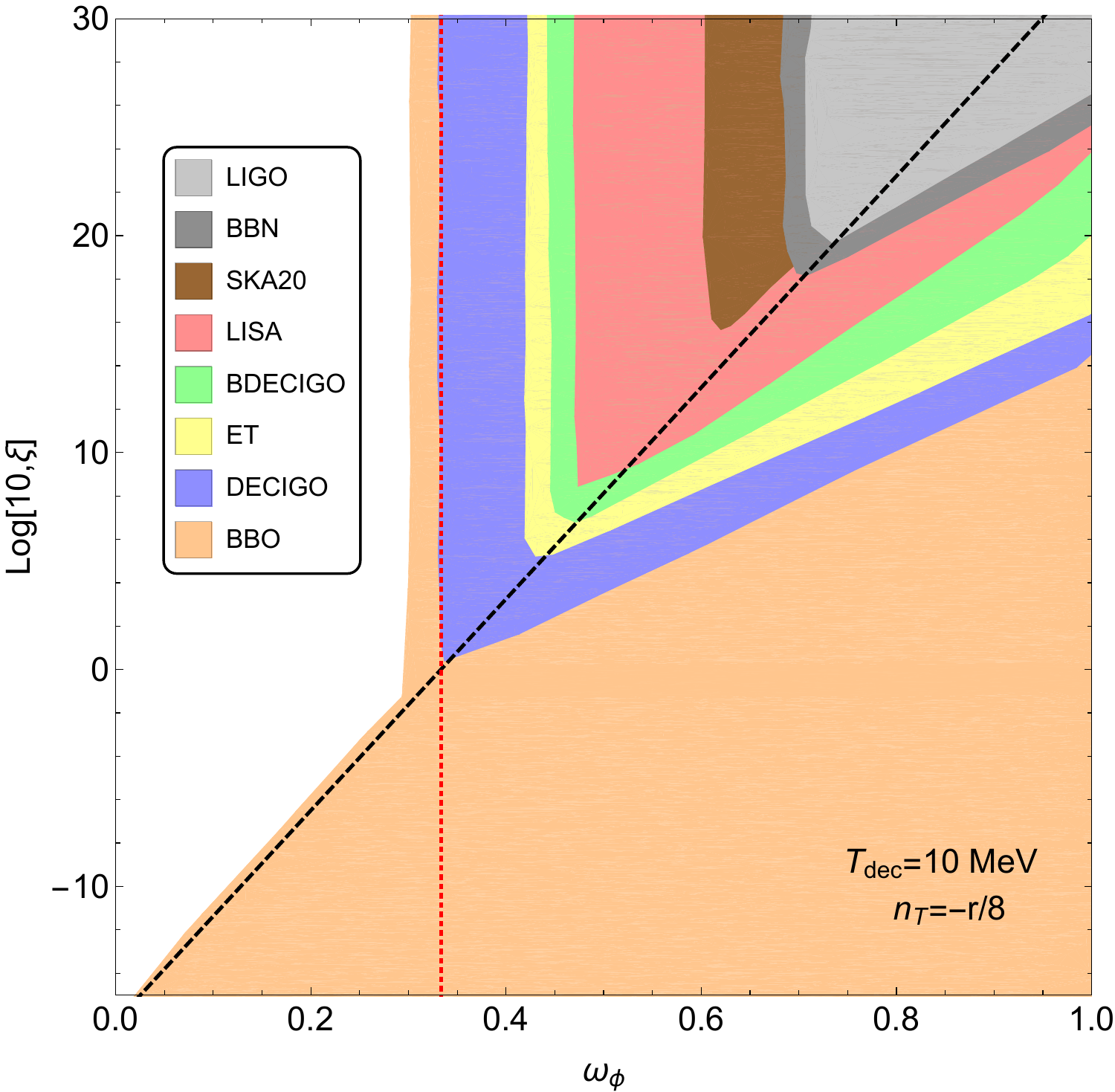}
\includegraphics[height=0.46\textwidth]{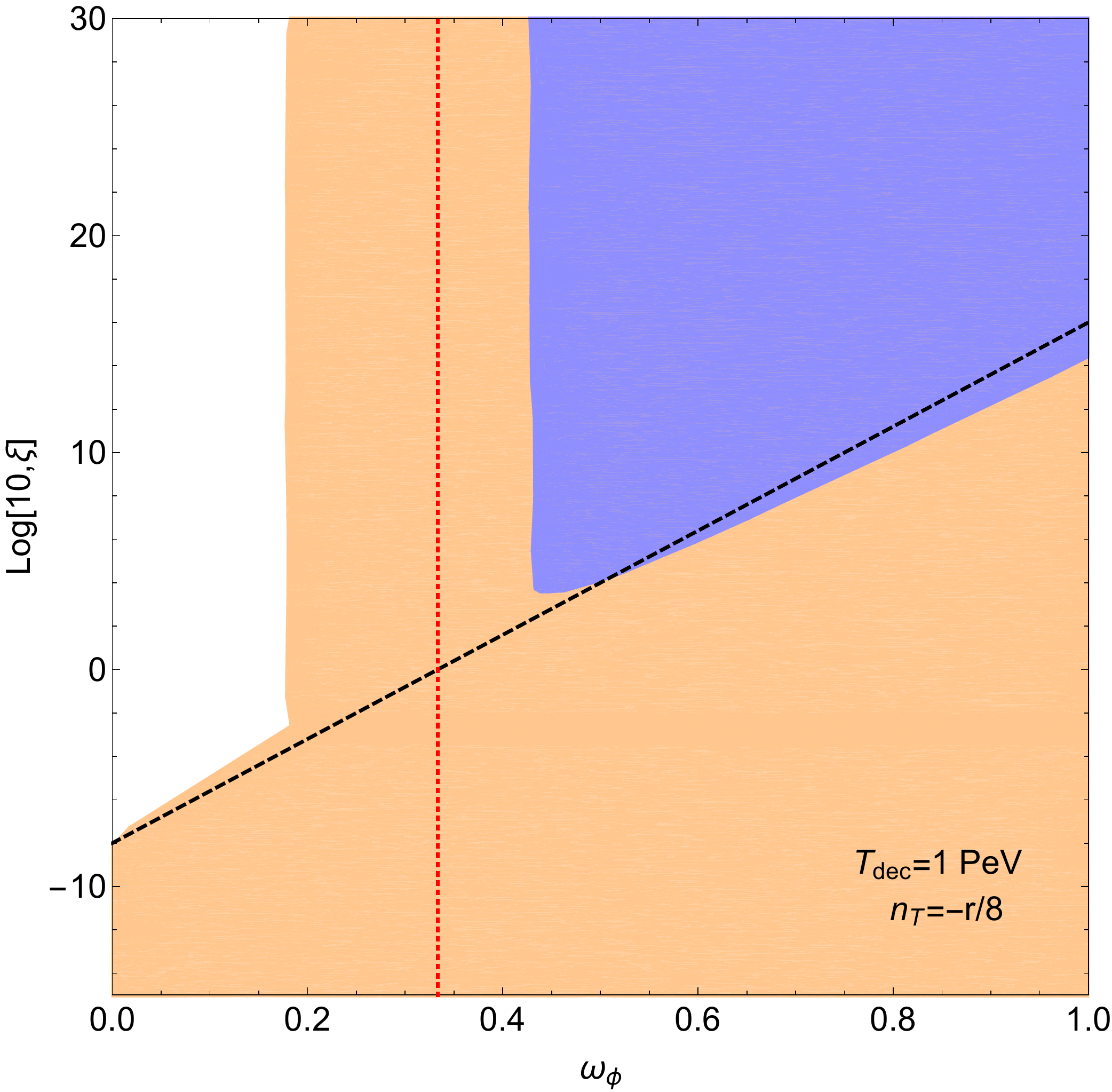}
\caption{Regions of the parameter space that could be tested by different observatories assuming the consistency relation, eq.~\eqref{rnt}, and $r=0.07$, for $\Tdec=10$~MeV (left panel) and 1~PeV (right panel).
The black dashed lines correspond to $\xi=\ximin$. The dotted red lines show the case $\op=1/3$.
}
\label{fig:treh-nT1}
\end{center}
\end{figure}
\begin{figure}[t]
\begin{center}
\includegraphics[height=0.46\textwidth]{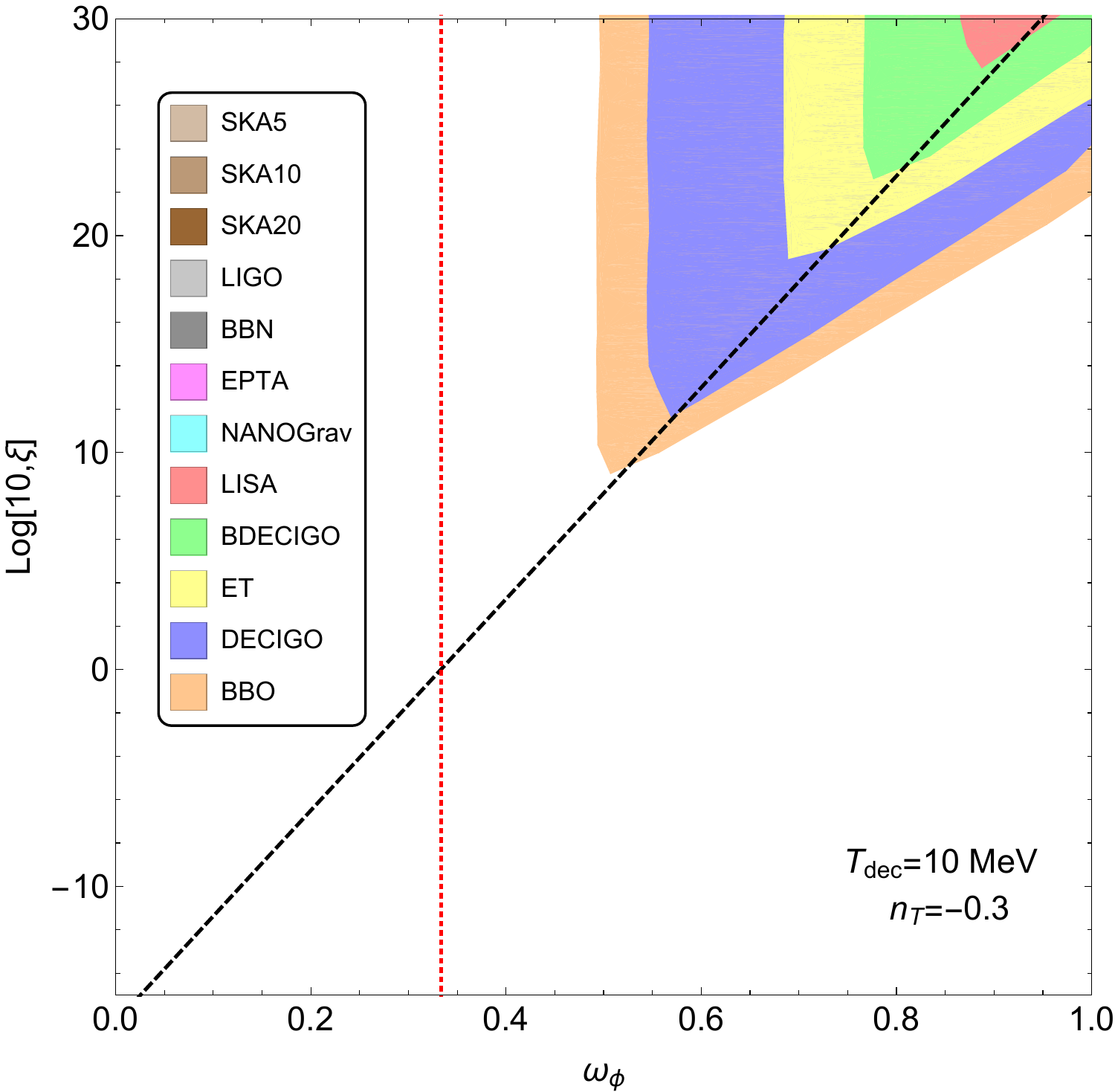}
\includegraphics[height=0.46\textwidth]{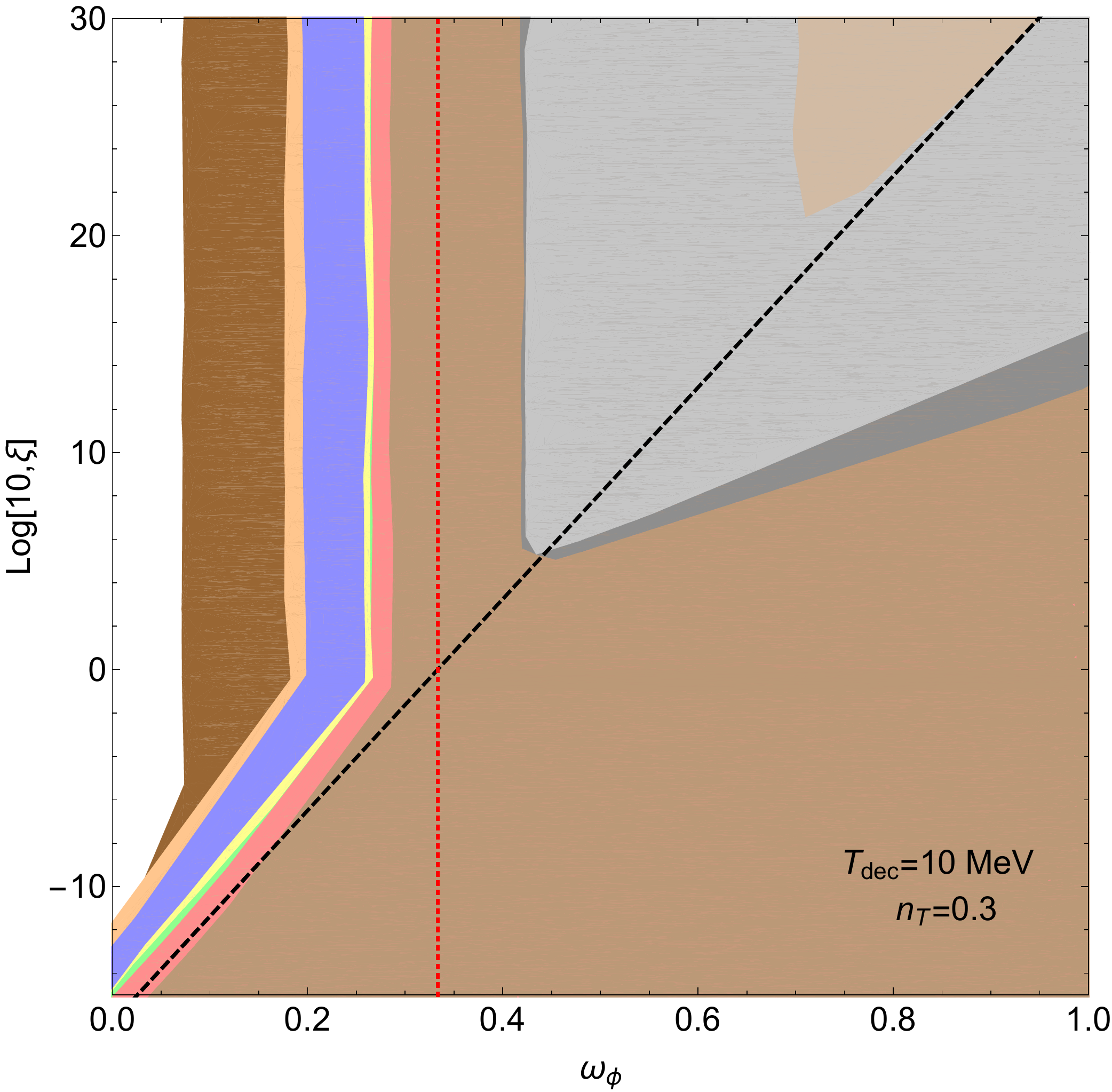}
\caption{Regions of the parameter space that could be tested by different observatories assuming $n_T=-0.3$ (left panel) and $n_T=0.3$ (right panel), for $\Tdec=10$~MeV.
The black dashed lines correspond to $\xi=\ximin$.
The red dotted lines show $\op=1/3$.}
\label{fig:treh-nT2}
\end{center}
\end{figure}

Figure~\ref{fig:treh-nT1} depicts the regions of the parameter space that could be probed by different observatories assuming the consistency relation in eq.~\eqref{rnt} and $r=0.07$, for $\Tdec=10$~MeV (left panel) and 1~PeV (right panel).
However, the consistency relation may not hold.
In fig.~\ref{fig:treh-nT2} it is shown the regions of the parameter space that could be probed by different observatories assuming $n_T=-0.3$ (left panel) and $n_T=0.3$ (right panel), for $\Tdec=10$~MeV.
The black dashed lines correspond to $\xi=\ximin$, the red dotted lines to $\op=1/3$.

The spectrum of PGW taking into account the possibility of a non-vanishing tensor tilt for modes which cross the horizon at scale factors in the range $\aeq\ll\ahc\ll\aend$ (similar to case \hyperref[sec:3]{2}) can be estimated to be
\begin{equation}
\OGW(\eta_0,\,k)\approx \frac{A_T\,\tilde{H}_{\text{max}}^{\frac{4}{1+3\op}}~\xi^{\frac{2}{1+3\op}}~\amax^{\frac{6(1+\op)}{1+3\op}}}{24\,a^4_0\,H^2_0\,{\tilde{k}}^{n_T}}\,k^{\frac{2(3\op-1)}{1+3\op}+n_T}.
\end{equation}
The extra $k^{n_T}$ dependence boosts (deteriorates) the detection prospects for the primordial GWs for $n_T>0$ ($n_T<0$), as shown in figs.~\ref{fig:treh-nT1} and~\ref{fig:treh-nT2}.
In particular, the right panel of fig.~\ref{fig:treh-nT2} shows a huge improvement on the detection possibilities in the case where $n_T=0.3$.

As done in the previous section the typical minimum value of the equation of state parameter $\op$ that can be probed by a given experiment happens in case~\hyperref[sec:2]{2}, when $\aeq\ll\ahc\ll\aend$ and $\xi>\ximin$.
\begin{equation}\label{appcons2}
    \omega_{\phi,\,\text{min}}\approx \frac43\frac{\ln \left[\left(\frac{\hs(T_0)}{\hs(\Tdec)}\right)^\frac13\left(\frac{\gs(\Tdec)}{\gs(\Tmax)}\right)^\frac14\frac{a_0\,\tHmax\,\Tdec\,T_0}{\Tmax^2}\right]}{\ln \left[\left(\frac{\hs(\Tdec)}{\hs(T_0)}\right)^\frac23\left(\frac{\gs(\Tdec)}{\gs(\Tmax)}\right)^\frac12\frac{24\,a_0^2\,H_0^2\,\Omega_{\text{min}}\,\Tdec^2}{A_T\, k_\text{min}^2\,T_0^2}\left(\frac{\tilde{k}}{k_{\text{min}}}\right)^{n_T}\right]}-\frac13\,,
\end{equation}
where $\Omega_{\text{min}}$ is defined below eq.~\eqref{appcons1}.
This relation  matches with the minimum values of $\op$ obtained in figs.~\ref{fig:treh-nT1} and~\ref{fig:treh-nT2} by precise numerical solutions, if one considers the numerical values for parameters and the values for $\Omega_{\text{min}}$ and $k_{\text{min}}$ from experimental constraints.
These minima can take values smaller than $1/3$ due to the extra dependence of eq.~\eqref{appcons2} on $n_T$, which shows some scenarios with a short period of early matter domination coming from small values of $\xi$ and $n_T>0$ that can be probed by future experiments.

In case~\hyperref[sec:3]{3}, taking into account the tilt of the primordial spectrum, the maximum $\xi$ then can be probed by a given experiment becomes
\begin{equation}\label{eq:ximaxomegaphiscale}
 \xi\approx\left[\frac{\gs(\Tdec)\Tdec^4}{\gs(\Tmax)\Tmax^4}\right]\left[\left(\frac{\hs(T_0)}{\hs(\Tdec)}\right)^\frac43\frac{A_T}{24\,\Omega_{\text{min}}}\left(\frac{k_{\text{min}}}{\tilde{k}}\right)^{n_T}\left(\frac{\tHmax}{H_0}\right)^2\frac{T_0^4}{\Tdec^4}\frac{\gs(\Tmax)}{\gs(\Tdec)}\right]^{\frac{3(1+\op)}{4}}.
\end{equation}
Finally, the lower limit on $\xi$ in the case~\hyperref[sec:4]{4} becomes
\begin{equation}\label{eq:xiomegaphiscale}
\xi\approx\left[\frac{k_{\text{min}}}{\tHmax\,\amax}\right]^2\left[24\frac{a_0^4\,H_0^2\,\Omega_{\text{min}}}{A_T\,k_\text{min}^2\,\amax^2}\left(\frac{\tilde{k}}{k_{\text{min}}}\right)^{n_T}\right]^{\frac{1+3\op}{2}}.
\end{equation}
Equations~\eqref{appcons2}, \eqref{eq:ximaxomegaphiscale} and~\eqref{eq:xiomegaphiscale} allow to analytically understand the behaviors of the sensitivity curves in figs.~\ref{fig:treh-nT1} and~\ref{fig:treh-nT2}.

\section{Summary and Conclusions}
\label{sec:conclusion}
Inflation, as a well-motivated theory to explain the early Universe cosmological problems, predicts the existence of a primordial gravitational wave (PGW) background.
The spectrum of the inflationary gravitational waves (GW) depends on the power spectrum of primordial tensor perturbations generated during inflation, and the expansion rate of the Universe from the end of inflation until today. 
This stochastic GW background could be probed by different gravitational waves observatories. 
In this paper we studied the PGW spectrum in scenarios beyond the standard cosmological framework, where the evolution of the Universe is dominated by SM radiation.
In fact, we analyzed non-standard scenarios dominated by a long lived component with a general equation of state. 
These cases are common in several UV-complete beyond the SM theories.

First we revisited the PGW spectrum in the case of a standard cosmology (i.e. with an energy density dominated by SM radiation), taking particularly care of the  evolution of the relativistic degrees of freedom, fig.~\ref{fig:OGWSM}.
Then we present the formalism used in order to define the non-standard cosmology.
We assume that for some period in the early Universe, the total energy density was dominated by a component $\phi$ with a general equation of state parameter $\op$.
We also assume that this component decays solely into SM radiation.
In addition to $\op$, the non-standard cosmology was parameterized by the ratio $\xi$ of $\phi$ to SM radiation energy densities and the temperature $\Tend$ at which $\phi$ decays.
This framework completely fixes the evolution of the energy densities, the Hubble expansion rate of the Universe and the evolution of the photon temperature, and allow us to numerically track in detail their behavior, fig.~\ref{fig:rho}.

In section~\ref{sec:pgw-non}, PGW in non-standard cosmologies were studied.
In particular, in section~\ref{sec:classification} we have analyzed the different possibilities and phenomenological consequences in which non-standard cosmologies can impact the PGW spectrum.
This strongly depends on the moment where the perturbations cross the horizon, with respect to the different characteristic scales $\aeq$, $\astart$ and $\aend$.
These analytical results where confronted and validated with numerical computations, e.g. fig.~\ref{fig:OGW}.

Once a signal from PGW is found, GW experiments can start probing the equation of state of the early Universe, in a given inflationary scenario.
Using the projected limits from future GW detectors, we study the possibilities to probe the parameter space $[\op,\,\xi]$ in fig.~\ref{fig:treh-vinf}.
We explore the impact of varying the scale of inflation and the temperature at which $\phi$ decays in the case where the primordial GW spectrum is scale invariant.
The general case where the primordial GW spectrum has a scale dependence was also analyzed and the results were shown in fig.~\ref{fig:treh-nT1} (assuming the single-field slow-roll consistency relation) and fig.~\ref{fig:treh-nT2} (general case).
Additionally, we scanned over the parameter space $[r,\,n_T]$ in fig.~\ref{fig:r-nT} to find the minimum value of the scalar-to-tensor ratio that different GW experiments can probe.

\acknowledgments
We would like to thank Juan Pablo Beltrán-Almeida, Mohammad Ali Gorji, Toby Opferkuch, Javier Rubio, Ken'ichi Saikawa, Jürgen Schaffner-Bielich and Dominik Schwarz for valuable discussions.
NB is partially supported by Spanish MINECO under Grant FPA2017-84543-P. 
FH is supported by the Deutsche Forschungsgemeinschaft (DFG, German Research Foundation) - Project number 315477589 - TRR 211.
This project has received funding from the European Union's Horizon 2020 research and innovation programme under the Marie Sklodowska-Curie grant agreements 674896 and 690575, and from Universidad Antonio Na\-ri\-ño grants 2017239 and 2018204.

\appendix
\section{Appendix}\label{sec:appendix}
From eq.~\eqref{eq:T} one has that the temperature scales like 
\begin{eqnarray}
\left[\frac{\gs(\Tmax)}{\gs(\Tstart)}\right]^\frac14\,\frac{\Tmax}{\Tstart}&=&\frac{\astart}{\amax},\label{eq:T1}\\
\left[\frac{\gs(\Tdec)}{\gs(\Tstart)}\right]^\frac14\,\frac{\Tdec}{\Tstart}&=&\left(\frac{\astart}{\adec}\right)^{\frac38(1+\op)},\label{eq:T2}\\
\left[\frac{\hs(\Tdec)}{\hs(T_0)}\right]^\frac13\,\frac{\Tdec}{T_0}&=&\frac{a_0}{\adec}.\label{eq:T3}
\end{eqnarray}
Additionally, in the sudden decay approximation of $\phi$, the conservation of energy density implies
\begin{equation}
    \rR(T_1)+\rp(T_1)=\rR(T_2),
\end{equation}
and therefore
\begin{equation}
   \gs(T_1)\,T_1^4+\xi\,\gs(\Tmax)\,\Tmax^4\left[\left(\frac{\gs(T_1)}{\gs(\Tmax)}\right)^\frac14\frac{T_1}{\Tmax}\right]^{3(1+\op)}=\gs(T_2)\,T_2^4,
\end{equation}
\begin{equation}\label{temp12max}
    \gs(T_1)^{1/4}\,T_1=\left[\frac{1}{\xi}\,\left(\gs(T_2)\,T_2^4\right)\,\left(\gs(\Tmax)^{1/4}\,\Tmax\right)^{3\op-1}\right]^\frac{1}{3(1+\op)},
\end{equation}
where $T_1$ and $T_2$ are the temperatures just before and just after $\phi$ decays, respectively.
Taking into account the scaling of $\rp$ and that $\rp(\Tmax)=\xi\,\rR(\Tmax)$, one gets that
\begin{equation}\label{eq:T2T1}
    \left[\frac{\gs(T_2)}{\gs(T_1)}\right]^\frac14\,\frac{T_2}{T_1}=\left[\frac{\gs(\Tdec)}{\gs(\Tmax)}\right]^\frac14\frac{\Tdec}{\Tmax}\,\frac{\adec}{\amax}=\left[\xi\left(\left[\frac{\gs(T_2)}{\gs(\Tmax)}\right]^\frac14\frac{T_2}{\Tmax}\right)^{3\op-1}\right]^\frac{1}{3(1+\op)}.
\end{equation}
In this approximation, $T_2$ can be identified with $\Tdec$.
Equations~\eqref{eq:T1}, \eqref{eq:T2} and~\eqref{eq:T2T1} can be rewritten as
\begin{eqnarray}
\adec&=&\amax\left[\frac{\gs(\Tmax)}{\gs(\Tdec)}\right]^\frac14\,\frac{\Tmax}{\Tdec}\,\left[\xi\left(\left[\frac{\gs(\Tmax)}{\gs(\Tdec)}\right]^\frac14\frac{\Tmax}{\Tdec}\right)^{1-3\op}\right]^\frac{1}{3(1+\op)},\label{eq:aend}\\
\Tstart&=&\Tdec\,\left[\frac{\gs(\Tdec)}{\gs(\Tstart)}\right]^\frac14\,\left[\xi\left(\left[\frac{\gs(\Tmax)}{\gs(\Tdec)}\right]^\frac14\,\frac{\Tmax}{\Tdec}\right)^{1-3\op}\right]^\frac{1}{5-3\op},\\
\astart&=&\amax\,\xi^\frac{1}{3\op-5}\,\left[\left(\frac{\gs(\Tmax)}{\gs(\Tdec)}\right)^\frac14\,\frac{\Tmax}{\Tdec}\right]^\frac{4}{5-3\op}.
\end{eqnarray}
Additionally, the equality $\rR=\rp$ happens at
\begin{eqnarray}
\Teq&=&\left[\frac{\gs(\Tmax)}{\gs(\Teq)}\right]^\frac14\Tmax\,\xi^\frac{1}{1-3\op},\\
\aeq&=&\amax\,\xi^\frac{1}{3\op-1}.
\end{eqnarray}
Moreover, $\amax$ can be extracted from eqs.~\eqref{eq:T3} and~\eqref{eq:aend}, and has the form of eq.~\eqref{eq:amax}.

Finally, assuming $T_1,\,T_2\rightarrow \Tdec$ and using eq.~\eqref{temp12max}, the minimum value of $\xi$ that leads to a $\phi$ domination phase which affects the evolution of radiation energy density can be obtained as
\begin{equation}\label{eq:minxsi1}
\ximin\approx\left[\left(\frac{g_\star(\Tmax)}{g_\star(\Tdec)}\right)^\frac14\frac{\Tmax}{\Tdec}\right]^{3\op-1}\,.
\end{equation}
If $\xi\gg\ximin$, $\phi$ dominates for some period the expansion rate of the Universe and also modifies the radiation energy density as $\rho_{R}\sim a^{-\frac{3}{2}(1+\op)}$ until $\Tend\approx\Tdec$.
In the opposite case, if $\xi\ll\ximin$ for $\op<1/3$ the $\phi$ domination regime never happens. 
However, for $\op>1/3$ if $1<\xi\ll\ximin$ the Universe is dominated by $\phi$ but the radiation energy density is not significantly modified by the presence of $\phi$.


\bibliography{biblio}

\end{document}